\documentclass[aip,reprint]{revtex4-1}

\usepackage{amssymb}
\usepackage{amsmath}
\usepackage{booktabs}
\usepackage[ruled,noline]{algorithm2e}
\usepackage{subcaption}
\usepackage{multirow}
\usepackage{graphicx}
\usepackage{placeins}

\newcommand{\bt}[1]{\mathbf{#1}}
\newcommand{\best}[1]{\underline{\textbf{#1}}}

\usepackage{lineno}
\usepackage{cancel}
\usepackage{xcolor}
\definecolor{ntcol}{rgb}{0.0, 0.62, 0.24}

\begin{document}


\title{Guiding diffusion models to reconstruct flow fields from sparse data} 



\author{Marc Amorós-Trepat}
\email[]{marc.amoros@tum.de}
\author{Luis Medrano-Navarro}
\email[]{luis.medrano@tum.de}
\author{Qiang Liu}
\email[]{qiang7.liu@tum.de}
\author{Luca Guastoni}
\email[]{luca.guastoni@tum.de}
\author{Nils Thuerey}
\email[]{nils.thuerey@tum.de}
\affiliation{School of Computation, Information, and Technology, Technical University of Munich, Garching 85748, Germany}


\date{\today}

\begin{abstract}
The reconstruction of unsteady flow fields from limited measurements is a challenging and crucial task for many engineering applications. Machine learning models are gaining popularity for solving this problem due to their ability to learn complex patterns from data and to generalize across diverse conditions. Among these, diffusion models have emerged as being particularly powerful for generative tasks, producing high-quality samples by iteratively refining noisy inputs. In contrast to other methods, these generative models are capable of reconstructing the smallest scales of the fluid spectrum. In this work, we introduce a novel sampling method for diffusion models that enables the reconstruction of high-fidelity samples by guiding the reverse process using the available sparse data. Moreover, we enhance the reconstructions with available physics knowledge using a conflict-free update method during training. To evaluate the effectiveness of our method, we conduct experiments on 2 and 3-dimensional turbulent flow data. 
Our method consistently outperforms other diffusion-based methods in predicting the fluid's structure and in pixel-wise accuracy. This study underscores the remarkable potential of diffusion models in reconstructing flow field data, paving the way for leveraging them in fluid dynamics research and applications ranging from super-resolution to reconstructions of experiments.
\end{abstract}


\maketitle 

\section{Introduction}
\label{Introduction}

Fluid dynamics plays an essential role in many modern engineering applications, including vehicle and turbine design \cite{pfeiffer2016closed, strom2017intracycle, maurya2018characteristics}, active flow control \cite{Manohar2018sensorPlacement, brunton2015closed, sipp2016linear}, cardiac blood flow modelling \cite{yakhot2007reconstruction, sankaran2012patient}, ship wake detection \cite{graziano2016ship} and climate predictions \cite{kalnay2003atmospheric}.
Numerical simulations and experimental studies are key methods of investigating complex flow phenomena. However, both approaches have inherent limitations. Although advancements in computational fluid dynamics (CFD) and hardware capabilities now enable high-fidelity simulations \cite{slotnick2014cfd}, accurate simulation of large-scale industrial problems still requires significant computational resources \cite{carlberg2019recovering}. Thus, reconstructing fine flow structures from sparse data is still a critical topic for fluid simulations. Meanwhile, experimental studies are often limited by sensor numbers in the flow field, and evaluating complex flow behaviour through sparse signals from sensors is a huge challenge for experimental fluid dynamics \cite{qin2022fine, raffel2018particle}. 

Sparse reconstruction is a technique employed to recover or approximate the full-scale features of a system using only a limited subset of data, such as selected measurements, pixels, or observations. It has been pivotal in aerodynamic applications, as seen in early work of Bui-Thanh et al. \cite{bui2004aerodynamic}, where they utilized Proper Orthogonal Decomposition (POD) to resolve complex flow fields and aerodynamic forces efficiently. Modifications of this method like Gappy-POD \cite{venturi2004gappy, raben2012adaptive} or other linear methods such as Dynamic Mode Decomposition (DMD) \cite{schmid2010dynamic, tu2013dynamic} have been used to extract the coherent structures and correlation features of a fluid flow, which can later be used to reconstruct its high-resolution representation. A different approach was used in the work of Callaham et al. \cite{callaham2018robust}, where sparse representation was leveraged to reconstruct target states from a dictionary of precomputed examples, showing better robustness than the previously introduced methods when encountering noisy data. Loiseau et al. \cite{loiseau2018sparse} presented a sparse reduced-order modeling framework that combined feature extraction, sparse nonlinear dynamics identification, and modal decomposition. They used it to reconstruct full-state fluid dynamics from typical experimental data such as time-resolved sensor data and optionally also non-time-resolved Particle Image Velocimetry (PIV) snapshots. The article from Zaki et al. \cite{zaki2021limited} proposed a hybrid approach bridging physics-based modelling and data-driven optimization, where an adjoint-variational method is introduced to reconstruct turbulent flow.

Meanwhile, the success of machine learning methods has motivated several studies addressing the challenges in data reconstruction with established machine learning pipelines. Deng et al. \cite{deng2019time} tried reconstructing flows using discrete point measurements and non-time-resolved PIV measurements based on long short-term memory (LSTM) networks. In the work of Erichson et al. \cite{erichson2020shallow}, Shallow Neural Networks (SNN) were explored to reconstruct a flow field from sparse sensor measurements. This same architecture was leveraged to reconstruct a flow over a stalled aerofoil in the work of Carter et al. \cite{carter2021data}, combining it with linear state estimation methods based on classical compressed sensing and extended POD methodologies. In this same field of aerodynamics, Zhong et al. \cite{zhong2023sparse} used a Convolutional Neural Network (CNN) combined with the multi-layer perceptron (MLP) to reconstruct the vorticity field surrounding the airfoil. Dubois et al. \cite{dubois2022machine} compared several machine learning techniques for flow data reconstructions with limited measurements, demonstrating the potential of deep learning methods in the rapid estimation of complex flows.

Solving the sparse reconstruction problem can also be seen as a super-resolution task, where our set of sparse sensor measurements compose a low-resolution sample, and we aim to reconstruct its high-resolution counterpart \cite{fukami2023super}. Sun et al. \cite{sun2020physics} proposed a physics-constrained Bayesian neural network (PC-BNN) to reconstruct a flow field, modelling their problem as a super-resolution task. Fukami et al. \cite{fukami2021global} took this same approach by computing a Voronoi tessellation of the sparse information and then using it as an input for a CNN, an established architecture for image super-resolution \cite{dong2015image}. Another example is the work of Güemes et al. \cite{guemes2022super}, where they presented a framework called RaSeedGAN that interpolated the scattered data onto a regular grid and afterwards upsampled it using a Generative Adversarial Network (GAN).

These methods have demonstrated remarkable performance in reconstructing fluid flows within the specific scenarios and parameter ranges they were designed to address. However, their effectiveness often diminishes when applied to cases that differ from their training data, such as flows with significantly different Reynolds numbers or unseen flow configurations. As a result, there is a need for approaches that are accurate within their domain and robust across a wide range of flow conditions and problem settings.

In recent years, diffusion models \cite{sohl2015deep,song2020score,ho2020denoising} have shown superior performance in tasks like image generation \cite{dhariwal2021diffusion}, image editing \cite{meng2021sdedit, avrahami2022blended}, image inpainting \cite{lugmayr2022repaint, chung2022come}, and image colorization \cite{saharia2022palette, yue2024resshift}. Following this success, there has been a rising interest in solving image super-resolution tasks with these models to improve in reconstruction accuracy. 
A direct approach to achieve this goal is to give the low-resolution image 
as an input to the model, and train it to map the high-resolution counterpart \cite{saharia2022image, rombach2022high}. Other approaches utilize a pre-trained diffusion model and only modify the sampling methods \cite{chung2022come, choi2021ilvr, yue2024difface, wang2024exploiting}. More recently, diffusion models have also been used to predict physical flows. Liu et al. \cite{liu2024uncertainty} explored their use as uncertainty-aware surrogate models for turbulence simulations, demonstrating their ability to capture full solution distributions and outperform traditional probabilistic approaches. Moreover, Shu et al. \cite{shu2023physics} presented a super-resolution method for reconstructing high-fidelity flow fields using this same architecture. Diffusion models have also been used to predict nonlinear fluid fields from discrete simulation data \cite{yang2023denoising}, turbulent rotating flows \cite{li2023multi}, and supersonic flows \cite{abaidi2025exploring}. With their generative capabilities and the proficiency to accurately learn complex distributions, diffusion models emerged as a promising candidate to enhance the existing reconstruction methods found in the literature. 
They are especially interesting in combination with physical priors, such as conservation laws.
These methods can also be adapted to solve inverse problems, similar to our reconstruction problem. One example is Diffusion Posterior Sampling (DPS)~\cite{chung2022diffusion}, which modifies the score function of diffusion models to make the sample closer to the observation during inference. This can lead to better accuracy, with the caveat of higher computational cost at inference time, as backpropagation through the network has to be computed in every step of the generation process. Furthermore, DPS only enforces the observed values as a soft constraint by steering the probability flow according to the observations. In our algorithm, we seek to address these limitations, avoiding backpropagation during the generation and providing a stronger constraint over the observations.

In terms of model architectures, early approaches primarily relied on CNNs applied to Cartesian grids \cite{airfoil}. Later, neural operators emerged, most notably in the form of Fourier Neural Operators (FNO) \cite{fno}, and Graph Neural Networks (GNNs) \cite{gnn_original, gnnlino}. 
More recently, spatial transformer architectures were employed as neural operators, e.g., via 
OFormer \cite{litransformer},
Universal Physics Transformers (UPT) \cite{upt} and Transolver \cite{transolver}, 
among others. Transformers have experienced particular interest
due to their ability to capture global and multi-scale interactions. Their main drawback is the computational   $N^2$ scaling of their inherent attention mechanism, where N is the number of grid nodes. Thus, vanilla attention can be prohibitive for large 2D or 3D problems. The P3D model~\cite{holzschuh2025p3dscalableneuralsurrogates}, which we employ in our experiments, employs shifted-window attention \cite{liu2021swin} to improve on this behavior.
Even with architectural advances, the computational cost of handling high-resolution 3D fields has restricted applications to low-resolution 3D settings with limited accuracy. This includes deterministic models \cite{super_3d_det_1,super_3d_det_2}, GANs \cite{super_3d_gan_1,super_3d_gan_2}, and diffusion approaches \cite{super_3d_diff_1,super_3d_diff_2}. In this context, our work advances diffusion-based super-resolution methods to 3D fields with high resolutions.

In the following, we build on state-of-the-art diffusion modeling techniques to develop a novel algorithm for sparse data reconstruction tasks of fluids that outperforms existing methods. Specifically, we first introduce a masking approach that improves the denoising step of diffusion models. This masking is combined with a modified, conflict-free update step to ensure that disagreements in the learning feedback from denoising and physics constraints do not hinder learning. We will demonstrate the gains in distributional accuracy and obeying physics constraints for a series of challenging fluid sparse data reconstruction tasks with turbulent flows.

\section{Methodology}
In this section, we will briefly introduce diffusion models to establish the theoretical framework required to present our new sampling method. Furthermore, we discuss the approaches used to train the underlying deep neural networks to ensure physical consistency.

\subsection{Diffusion models}\label{sec:DDPM}
In this work, we aim to reconstruct a high-fidelity sample $\bt{x_0}\in\mathbb{R}^{n \times n}$ from a subset of sparse true values $\bt{x_s}$ with denoising diffusion probabilistic models (DDPMs)~\cite{sohl2015deep,ho2020denoising}.
DDPMs are generative models that learn to gradually transform samples from a simple known distribution $p(\bt{x}_T)$ into samples from the target distribution $p(\bt{x}_0)$ that we want to model. Typically,  $p(\bt{x}_T)$ is a normal distribution $\bt{x}_T \sim \mathcal{N}(\bt{0}, \bt{I})$. The transformation from $p(\bt{x}_T)$ to $p(\bt{x}_0)$, usually known as \textit{reverse process}, is an iterative process described by the learnable  distribution $p_{\theta}(\bt{x}_{0:T})$:
\begin{equation}
    p_{\theta}(\bt{x}_{0:T}) := p(\bt{x}_T)\prod_{t=1}^{T} p_{\theta}(\bt{x}_{t-1} | \bt{x}_t),
\end{equation}
\begin{equation}
    p_{\theta}(\bt{x}_{t-1} | \bt{x}_t) := \mathcal{N}(\bt{x}_{t-1};\mu_{\theta}(\bt{x}_t, t), \Sigma_{\theta}(\bt{x}_t, t)) \label{eq:reverse_step}
\end{equation}
where $\bt{x}_1, \dots, \bt{x}_T$ are latent variables with the same dimensionality as the samples of the target distribution that represent intermediate steps, and $\theta$ are the learnable parameters. This process can be seen as a Markovian chain, where a network learns Gaussian transitions $p_{\theta}(\bt{x}_{t-1} | \bt{x}_t)$ that predict the next latent variable $\bt{x}_{t-1}$ given $\bt{x}_{t}$ as input. 

To construct a trainable objective for $\theta$, we also need to introduce the 
\textit{forward process}, where noise is gradually added to a sample from our target distribution until it is indistinguishable from pure noise. It is defined as follows:
\begin{equation}
\begin{split}
    q(\bt{x}_{1:T} | \bt{x}_{0}) &:= \sum_{t=1}^{T} q(\bt{x}_{t} | \bt{x}_{t-1}) \\
    q(\bt{x}_{t} | \bt{x}_{t-1}) &:= \mathcal{N}(\bt{x}_{t};\sqrt{1 - \beta_t} \bt{x}_{t-1}, \beta_t \bt{I}) 
\end{split}
\end{equation}
where $\beta_1, \dots, \beta_T$ are coefficients that schedule the amount of noise to be added at each step. An interesting property of the forward process is that we can express any arbitrary step $\bt{x}_t$ using the coefficients $\alpha_t := 1 - \beta_t$ and $\bar{\alpha}_t := \prod_{s=1}^{t} \alpha_s$:
\begin{equation}
    q(\bt{x}_t | \bt{x}_0) = \mathcal{N}(\bt{x}_{t};\sqrt{\bar{\alpha}_t} \bt{x}_{0}, (1 - \bar{\alpha}_t) \bt{I}) 
\end{equation}
meaning that we can actually compute $\bt{x}_t$ without computing the entire chain:
\begin{equation}
    \bt{x}_t = \sqrt{\bar{\alpha}_t} \bt{x}_{0} + \sqrt{1 - \bar{\alpha}_t}\epsilon, \hspace{0.4cm} \text{where} \hspace{0.4cm} \epsilon \sim \mathcal{N}(\bt{0}, \bt{I}) 
    \label{eq:forward_xt}
\end{equation}

With the above forward procedure, Ho et al. \cite{ho2020denoising} proposed a simple loss function where the network directly learns the noise $\epsilon$ in the forward procedure: 
\begin{equation}
    L_{\text{simple}} := \mathbb{E}_{t, \bt{x}_0, \epsilon} \left[ \Vert \epsilon - \epsilon_{\theta} (\sqrt{\bar{\alpha}_t} \bt{x}_{0} + \sqrt{1 - \bar{\alpha}_t}\epsilon, t) \Vert^2   \right]
    \label{eq:lsimple}
\end{equation}
Using a fixed variance for the reverse process to $\Sigma(\bt{x}_t, t)~=~\frac{1-\bar{\alpha}_{t-1}}{1-\bar{\alpha}}\beta_t\bt{I}$, the reverse step becomes:
\begin{equation}
    \bt{x}_{t-1} = \tilde{\bt{x}}_0^{(t)} \sqrt{\bar{\alpha}_{t-1}}  + \epsilon_\theta\left(\bt{x}_t, t\right) \sqrt{1 - \bar{\alpha}_{t-1}} 
    \label{eq:xt-1}
\end{equation}
where $\tilde{\bt{x}}_0^{(t)}$ is the \textit{denoised observation}, a prediction of the generated sample $\bt{x}_0$ at the step $t$, which can be obtained by rearranging equation (\ref{eq:forward_xt}):
\begin{equation}
    \tilde{\bt{x}}_0^{(t)}:=\frac{\bt{x}_t-\sqrt{1-\bar{\alpha}_t} \cdot \epsilon_\theta\left(\bt{x}_t, t\right)}{\sqrt{\bar{\alpha}_t}}
    \label{eq:denoised_observation}
\end{equation}

\subsection{Masked Sampling of Diffusion Models}
\label{sec:meth_sampling}
The DDPM algorithm enables training a network $\epsilon_\theta$ that transforms Gaussian noise $\bt{x}_T$ to a high-fidelity flow field $\bt{x}_0$. In this section, we present a novel sampling method where the reverse process is further guided to generate a certain flow field corresponding to a given sparse observation. 

The idea behind our method is based on the key observation that the sparse inputs $\bt{x_s}$ are authentic data points that should be preserved in the final generated sample. Thus, we introduce the sparse information $\bt{x_s}$ into the denoised observation $\tilde{\bt{x}}_0^{(t)}$ by applying a masking operation. Specifically, we define a binary mask $m \in \mathbb{R}^{n \times n}$, where entries corresponding to known ground truth values are set to 1, while all other positions are assigned a value of 0. After computing the denoised observation $\tilde{\bt{x}}^{(t)}_0$, we include the available true values via:
\begin{equation}
    \tilde{\bt{x}}^{(t)}_m := \tilde{\bt{x}}^{(t)}_0 \cdot (1 - m_t) + \bt{x_s} \cdot m_t
\end{equation}
where all the operations are performed \textit{element-wise}. Since the prediction of $\bt{x}_0$ at the start of the reverse process is usually very poor, we introduce a scheduler $m_t~=~m~\cdot~\gamma_t$ where $\gamma_t \in [0, 1]$ to regulate the strength of the mask. $\gamma_t$ is stronger when the prediction is nearly random (t is close to T) and weaker when we are closer to the solution (t is close to 0). 

After masking the prediction with the real values, we can compute the next latent variable $\bt{x}_{t-1}$ following equation (\ref{eq:xt-1}), substituting the denoised observation by our masked version $ \tilde{\bt{x}}_m^{(t)}$:
\begin{equation}
    \bt{x}_{t-1} = \tilde{\bt{x}}_m^{(t)}\sqrt{\bar{\alpha}_{t-1}}
    + \epsilon_{\theta}(\bt{x}_t, t)\sqrt{1-\bar{\alpha}_{t-1}}
    \label{eq:mask_xt-1}
\end{equation}
Starting the reverse process from pure noise $\bt{x}_T$, equation (\ref{eq:mask_xt-1}) will guide the model towards predicting a high-fidelity flow field that corresponds to the sparse observation $\bt{x}_s$.

\begin{figure} [ht!]
  \centering
\begin{subfigure}{.5\linewidth}
  \centering
  \includegraphics[height=0.8\linewidth]{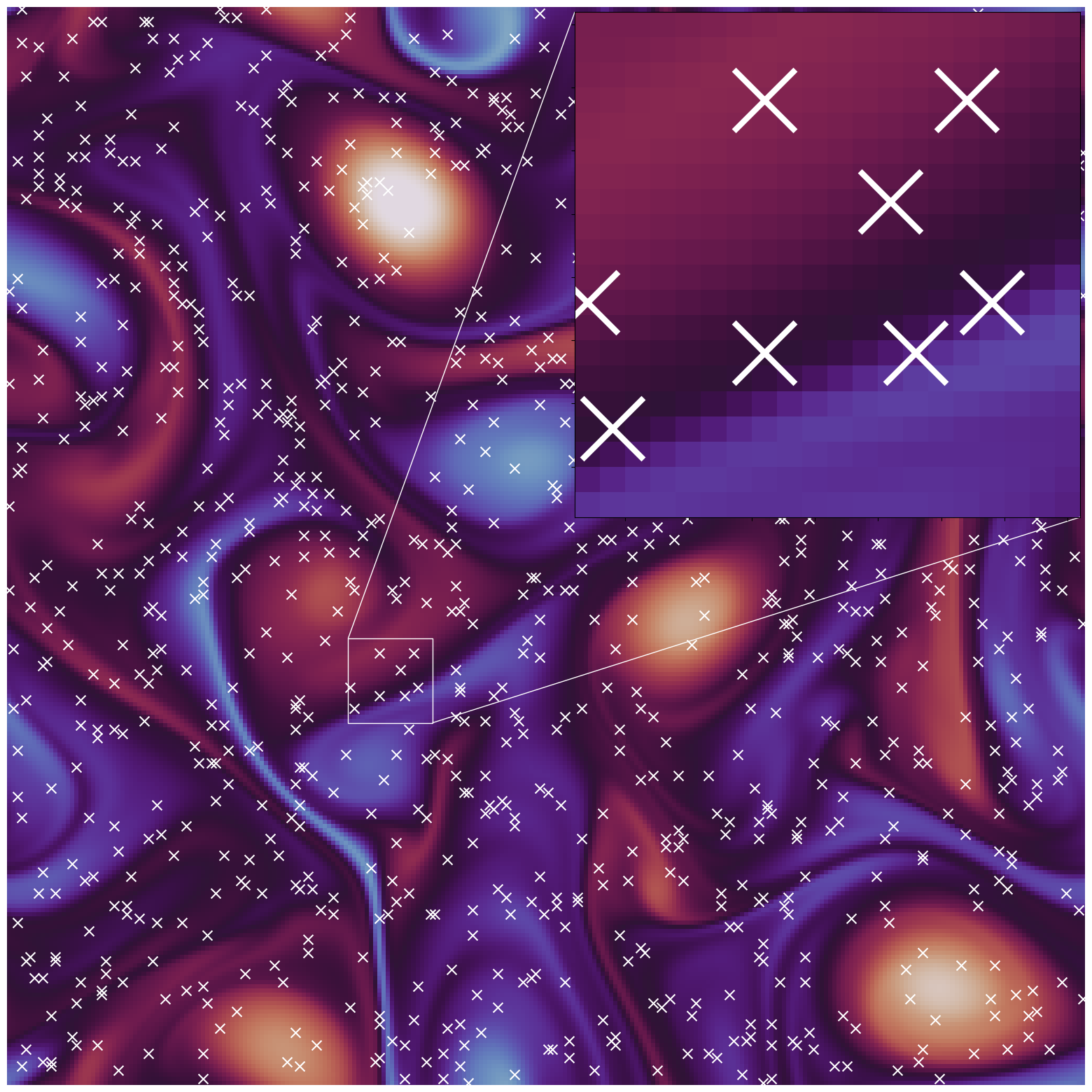}
  \caption{}
\label{fig:smooth:a}
\end{subfigure}%
\begin{subfigure}{.5\linewidth}
  \centering
  \includegraphics[height=0.8\linewidth]{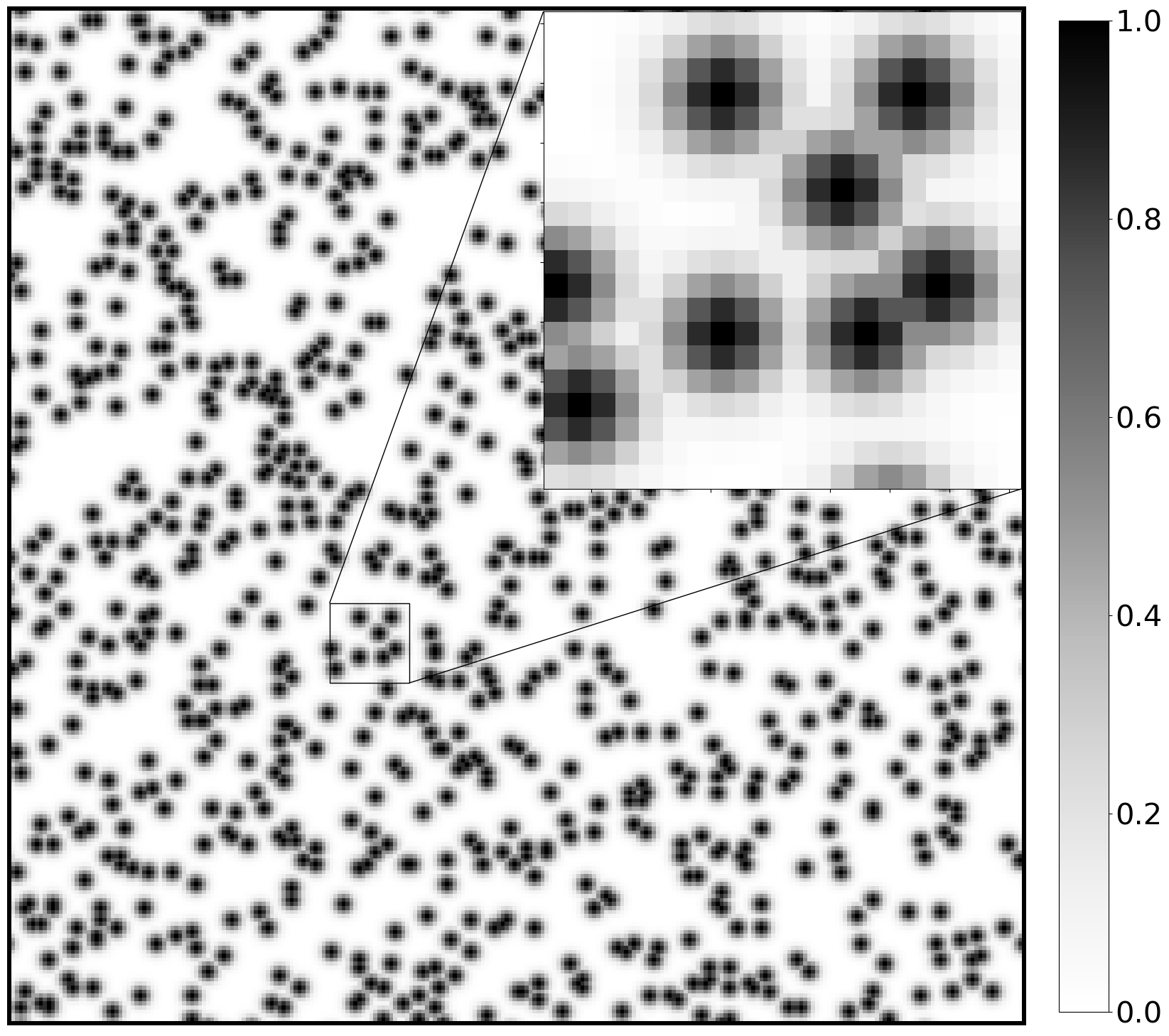}
  \caption{}
\label{fig:smooth:b}
\end{subfigure}
\caption{a) Sparse data $\bt{x}_s$ marked with white crosses over ground truth. b) Smooth Gaussian mask $m^s$.}
\label{fig:smooth}
\end{figure}

\begin{figure*}[ht!]
    \centering
    \includegraphics[width=0.8\textwidth]{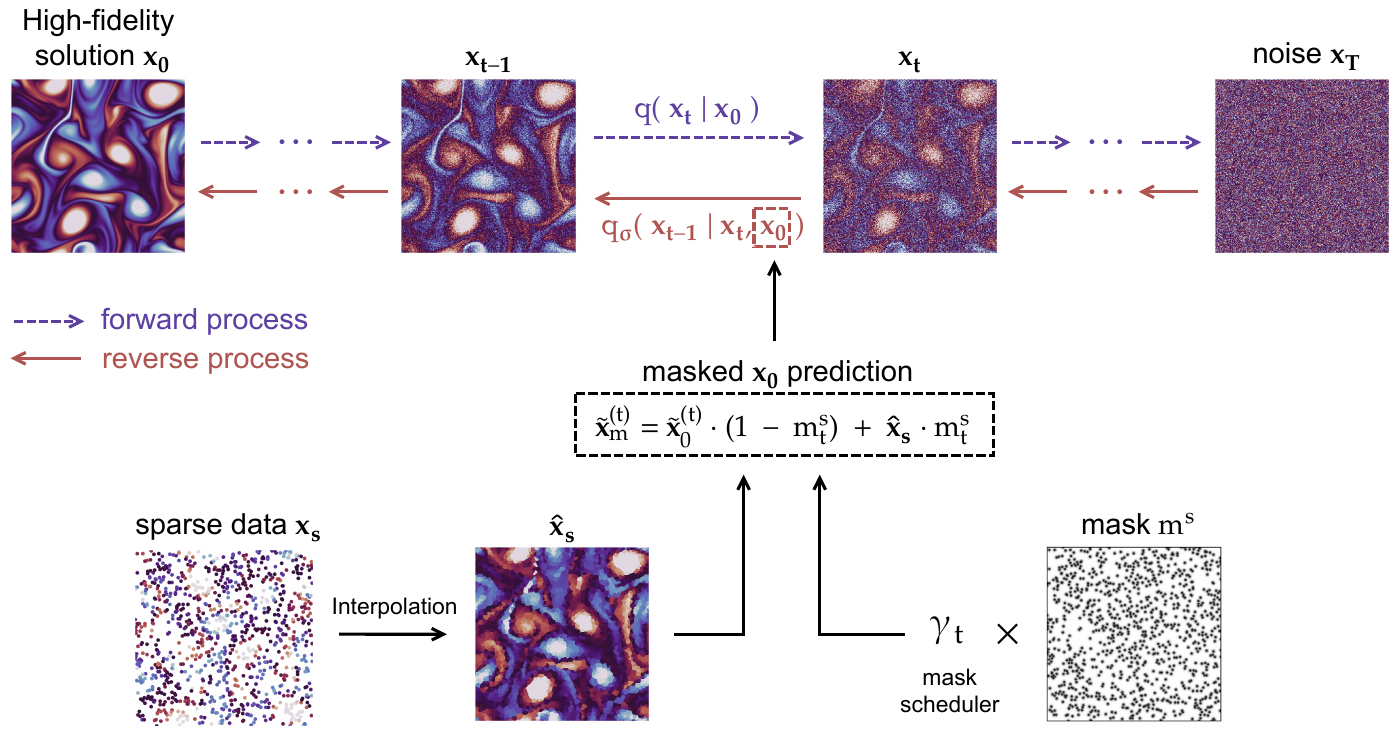}
    \caption{Sketch of our proposed masked diffusion method, where a smooth mask is applied to improve the prediction of $\bt{x}_0$ during the reverse process.}
    \label{fig:masking_method_sketch}
\end{figure*}

However, as described so far, the mask only yields single-point values onto the field, which can introduce discontinuities. Thus, we also propose a smooth Gaussian mask $m^s$. Given the set of sparse data points $P~=~\{\bt{p}^1, \dots, \bt{p}^K\}$ representing the input $\bt{x_s}$, we define a function $f_k(\bt{x})$ for each point $\bt{p}^k$ that returns the value of a Gaussian centred at that point with a smoothing factor $\sigma$:
\begin{equation}
    \begin{split}
        f_k(\bt{x}) &= \exp \left( - \frac{\Vert \bt{x} - \bt{p}^k\Vert_2}{2 \sigma^2} \right) \\
    \end{split}
\end{equation}
We define the  Gaussian operator $m^s = G_{\bt{s}}(\bt{x_s})$, which computes this smooth mask $m^s$ by evaluating each function $f_k(\bt{x})$ across all cells of the mask. An example of a resulting mask can be seen in Figure~\ref{fig:smooth:b}. Meanwhile, since we have no true information of the values surrounding each mask point $\bt{p}^k$, we compute the interpolation between the true points using an interpolation operator $\bt{\hat{x}_s}=U_{\bt{s}}(\bt{x_s})$. Now, we can also mask the prediction $\tilde{\bt{x}}^{(t)}_0$ with intermediate values between the true sparse data following:
\begin{equation}
    \tilde{\bt{x}}^{(t)}_m = \tilde{\bt{x}}^{(t)}_0 \cdot (1 - m_t^{s}) + \hat{\bt{x}}_{\bt{s}} \cdot m_t^{s}
\end{equation}
Similarly to the binary mask, we also apply a scheduler $m_t^s = m^s \gamma_t$ to the smooth mask. A comparison between the point-wise and the Gaussian masking methods can be found in Appendix \ref{app:masks}, which clearly displays the benefits of using the Gaussian mask. Thus, we continue with this masking method and we call it \textit{masked diffusion}. Figure~\ref{fig:masking_method_sketch} illustrates the method, and a detailed summary of the procedure can be found in Algorithm~\ref{alg:masking_method}.

\begin{algorithm}[ht!]
\caption{Masked diffusion algorithm} \label{alg:masking_method}

      \KwIn{Sparse data sample $\bt{x_s}$} 
     $m^s = G_s(\bt{x_s})$ \; 

	$\hat{\bt{x}}_{\bt{s}} = U_s(\bt{x}_{\bt{s}})$ \; 
	
	$\mathbf{x}_T \sim \mathcal{N}(0, \bt{I})$ \; 
	
	\For{$t = T, \dots, 1$}
	{        
		$\;\;\;\; m^s_t = m^s \cdot \gamma_t$
		
		$\;\;\;\; \tilde{\bt{x}}_0^{(t)} = \left(\bt{x}_t-\sqrt{1-\bar{\alpha}_t} \cdot \epsilon_{\theta}(\bt{x}_t, t)\right)/\sqrt{\bar{\alpha}_t}$ \; 
		
		$\;\;\;\;\tilde{\bt{x}}^{(t)}_m = \tilde{\bt{x}}^{(t)}_0 \cdot (1 - m_t^{s}) + \hat{\bt{x}}_{\bt{s}} \cdot m_t^{s}$ \; 
		
		$\;\;\;\;\bt{x}_{t-1} = \tilde{\bt{x}}_m^{(t)}\sqrt{\bar{\alpha}_{t-1}} + \epsilon_{\theta}(\bt{x}_t, t)\sqrt{1-\bar{\alpha}_{t-1}}$ \; 
	}
	
	\Return $\mathbf{x}_0$ \;

\end{algorithm}

\subsection{Physics-based Diffusion Training}
\label{sec:phy-train}
The high fidelity samples $\bt{x}_0$ usually satisfy physical constraints, i.e., $\mathcal{F}(\bt{x})=0$, where $\mathcal{F}$ could be an underlying PDE stemming from conservation laws or representing boundary conditions. To enhance the physical coherence of the samples generated by DDPMs, we can integrate an additional physical residual loss into the training as:
\begin{equation}
    L_{\text{phy}} := \mathbb{E}_{t, \bt{x}_0, \epsilon} \left[ \Vert \epsilon - \epsilon_{\theta} (\bt{x}_t, t) \Vert^2 + \lambda \Vert \mathcal{F}(\tilde{\bt{x}}_0^{(t)})\Vert_2  \right]
\end{equation}
where $\lambda$ is a balance factor for the physical residual.
Similar methods have been proposed by previous works \cite{bastek2024physics,shan2024pird}, such as the Physics-Informed Diffusion Models (PIDMs) \cite{bastek2024physics}. In PIDM, a central difficulty is that the scaling factor is highly dependent on the specificity of the training task at hand. We consider an alternative approach where the rescaling factor is dynamically adjusted to balance the importance of each term in the global loss. This balance is computed via the ratio of the noise step and the physics residual loss as:
\begin{equation}
\lambda_{\text{dyn}} =  \frac{\Vert \epsilon - \epsilon_{\theta} (\bt{x}_t, t) \Vert^2}{\Vert\mathcal{F}(\tilde{\bt{x}}_0^{(t)})\Vert_2}
\label{eq:lambda-dyn}
\end{equation}

To improve on these manually-scaled and heuristic approaches for balancing, we build on a recent method for conflict-free updates, \textit{ConFIG} \cite{liu2024config}, to automatically balance the loss terms. Defining $\boldsymbol{g}_D={\partial \Vert \epsilon - \epsilon_{\theta} (\bt{x}_t, t) \Vert_2}/{\partial \theta}$ and $\boldsymbol{g}_F={\partial \Vert \mathcal{F}(\tilde{\bt{x}}_0^{(t)})\Vert_2}/{\partial \theta}$ as the gradients for diffusion loss and physical residual loss terms, the ConFIG method returns a conflict-free update direction $\boldsymbol{g}_c$ as
\begin{equation}
\begin{split}
&\boldsymbol{g}_v=\mathcal{U}\left[\mathcal{U}\left(\mathcal{O}\left(\boldsymbol{g}_D, \boldsymbol{g}_F\right)\right)+\mathcal{U}\left(\mathcal{O}\left(\boldsymbol{g}_F, \boldsymbol{g}_D\right)\right)\right] \\
    &\boldsymbol{g}_c=\left(\boldsymbol{g}_D^{\top} \boldsymbol{g}_v+\boldsymbol{g}_F^{\top} \boldsymbol{g}_v\right) \boldsymbol{g}_v 
\end{split}
\label{eq:ConFIG}
\end{equation}
where $\mathcal{U}(\boldsymbol{g}_i) = \boldsymbol{g}_i / |\boldsymbol{g}_i| + \varepsilon$ is the normalizing operator and $\mathcal{O}(\boldsymbol{g}_i, \boldsymbol{g}_j) = \boldsymbol{g}_j - \frac{\boldsymbol{g}_i^{\top} \boldsymbol{g}_j}{|\boldsymbol{g}_i|^2}\boldsymbol{g}_i$ the orthogonality operator. 

Provably, $\boldsymbol{g}_D^\top \boldsymbol{g}_c >0$ and $\boldsymbol{g}_F^\top \boldsymbol{g}_c >0$ hold, meaning that optimization along $\boldsymbol{g}_c$ will decrease both the diffusion loss and physical residual loss simultaneously. The generalization of the operator to more than two gradients is presented in Liu et al.'s article \cite{liu2024config}.
As shown in equation (\ref{eq:ConFIG}), the magnitude of $\boldsymbol{g}_v$ in the ConFIG method is rescaled based on the sum of each loss-specific gradient's projection on the conflict-free direction. However, for sparse reconstruction tasks, the physical residual loss is usually significantly larger than the diffusion loss. This leads to the magnitude change of the $\boldsymbol{g}_v$ mainly tracking the magnitude change of the residual gradient. For momentum-based optimizers like Adam, this represents a problem, as these optimizers adaptively change the learning rate according to the variation of the gradient. In turn, it makes the balancing approach of the ConFIG method ineffective. To counteract this undesirable behavior, we propose the following modification to the ConFIG method, which computes the magnitude of the resulting gradient after normalizing both loss-specific gradients:
\begin{equation}
        \boldsymbol{g}_c=\left(\mathcal{U}(\boldsymbol{g}_D)^{\top} \boldsymbol{g}_v+\mathcal{U}(\boldsymbol{g}_F)^{\top} \boldsymbol{g}_v\right) \boldsymbol{g}_v 
\end{equation}
This results in a method that considers both loss terms equally when computing the gradient magnitude. At the same time, it also allows the Adam optimizer to compute the moments of the gradient in a balanced manner. The caveat of this modification is that we do not transfer any information about the original magnitude of $\boldsymbol{g}_D$ and $\boldsymbol{g}_F$ to the optimizer, as the new magnitude only contains the information of how similar the direction of $\boldsymbol{g}_v$ is to the directions of both loss terms. Nonetheless, our results show that the benefits of conflict-free learning updates clearly outweigh the lack of information about the original magnitudes.

\begin{figure} [ht!]
  \centering
\begin{subfigure}{\linewidth}
  \includegraphics[width=0.85\linewidth]{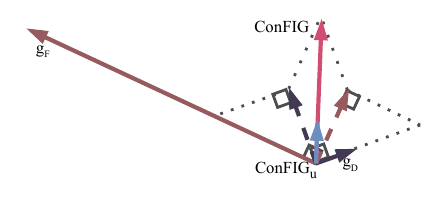}
  \caption{Conflicting gradients.}
\label{fig:config:a}
\end{subfigure}%
\hspace{1.3cm}
\begin{subfigure}{\linewidth}
  \includegraphics[width=0.7\linewidth]{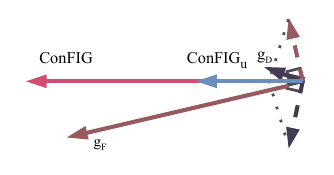}
  \caption{Similarly directed gradients.}
\label{fig:config:b}
\end{subfigure}
\caption{A comparison of ConFIG and $\text{ConFIG}_\text{u}$ methods, where $\boldsymbol{g}_F$ is the gradient of the physical loss and $\boldsymbol{g}_D$ the gradient of the diffusion loss.}
\label{fig:config_exp}
\end{figure}

We refer to the resulting method as $\text{ConFIG}_u$ in the following, and Figure~\ref{fig:config_exp} illustrates how our update strategy differs from the original ConFIG method for a specific two-dimensional example. The $\text{ConFIG}_\text{u}$ updates in blue preserve significantly more information from the diffusion updates and are generally more balanced. 
They effectively capture the agreement between the directions of the two loss gradients. We compare the performance of these training strategies in Section \ref{sec:train_meth}.

\section{Results and discussion}
In this section, we evaluate our method in two challenging scenarios:  2D Kolmogorov flow and 3D isotropic turbulence. We first compare the performance of different training methods, demonstrating how our $\text{ConFIG}_\text{u}$ training method outperforms existing approaches in generating physically consistent samples. Next, we investigate the effectiveness of our proposed masked diffusion sampling method for sparse data reconstruction. The results demonstrate how our method effectively reconstructs fine-grained details and preserves the structural integrity of the flow fields, outperforming baseline methods, particularly in capturing high-frequency components that are crucial in turbulent flows. Finally, we provide a qualitative and quantitative analysis of the reconstructed samples, including visualizations and an evaluation of the vorticity and Q-criterion distributions.

\subsection{Test cases}

\begin{figure*}[ht!]
    \centering
    \hspace{0.2cm}
    \begin{subfigure}{0.74\linewidth}
        \centering
        \includegraphics[width=0.8\linewidth]{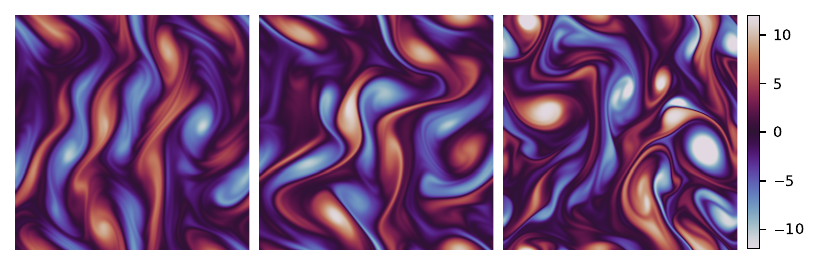}
        \caption{Examples of the 2D dataset, comprised of vorticity field snapshots of DNS simulations.}
        \label{fig:example_2D_data}
    \end{subfigure}
    
    \vspace{0.5cm} 

    \begin{subfigure}{0.7\linewidth}
        \centering
        \includegraphics[width=0.85\linewidth]{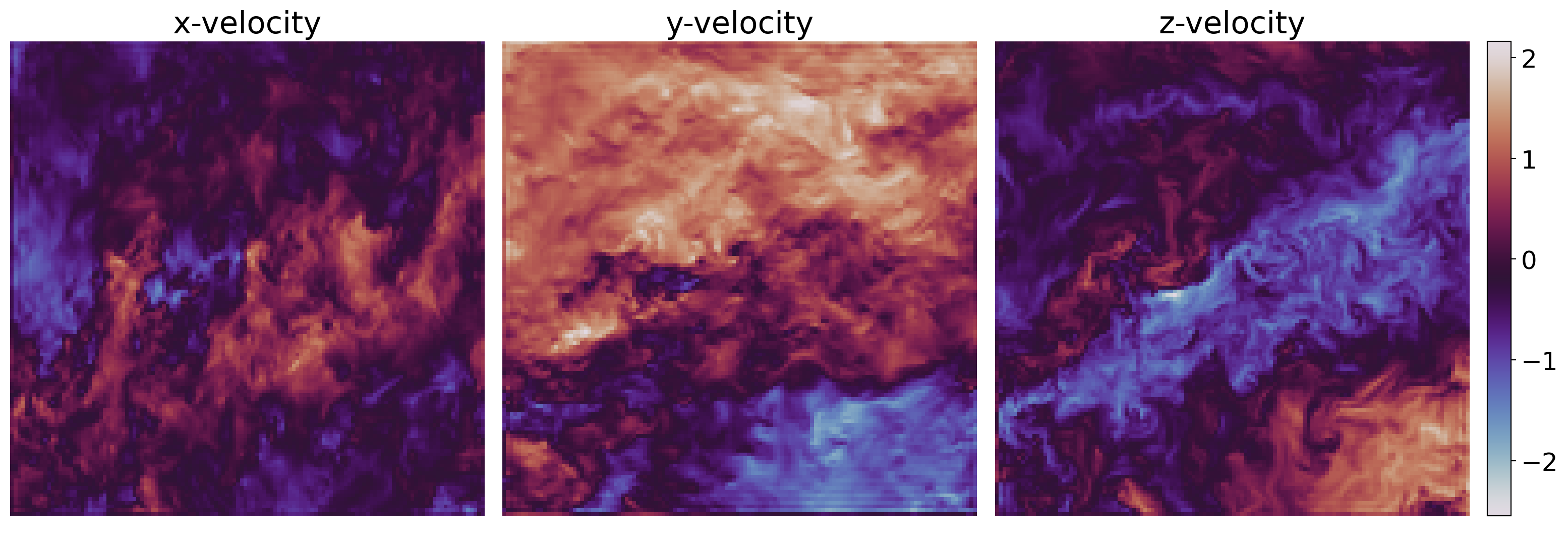}
        \caption{Example of one snapshot of the 3D dataset. We plot 2D z-slices through the middle of the domain for the three velocity components.}
        \label{fig:v_original}
    \end{subfigure}
    
    \caption{Datasets used: (a) 2D Kolmogorov dataset, (b) 3D isotropic turbulence.}
    \label{fig:datasets}
\end{figure*}

For the 2D Kolmogorov flow, we follow the configuration of the previous work \cite{shu2023physics}. The system is governed by the following Navier-Stokes equation in its vorticity form:
\begin{equation}
\frac{\partial \omega (\mathbf{x}, t)}{\partial t} + \mathbf{u}(\mathbf{x}, t) \cdot \nabla \omega (\mathbf{x}, t) =
    \frac{1}{Re} \nabla^2 \omega (\mathbf{x}, t) + f(\mathbf{x})
\label{eq:vorticity}
\end{equation}
where $\textbf{x} \in (0, 2\pi)^2$, $\omega$ is the vorticity, $\bt{u}$ the velocity field and $f(\bt{x})=-4\cos(4 x_2) - 0.1 \omega(\bt{x}, t)$ is the external force. The Reynolds number $Re$ is set to 1000. The simulation data contains 4240 samples of size $256 \times 256$, from which 3000 are randomly selected for training and the remaining 1240 to test the models. To evaluate the physical consistency of the generated samples, we compute the residual of equation (\ref{eq:vorticity}), where the time derivative is approximated with three generated consecutive timesteps as $\partial_t~\omega~\approx~(\omega(\bt{x}, t - \Delta t) + \omega(\bt{x}, t + \Delta t)) / (2\Delta t)$, and the spatial derivatives are calculated with the Fourier pseudo-spectral method. 

For the 3D turbulent scenario, we choose a forced isotropic turbulence simulation from the Johns Hopkins Turbulence Database (JHTDB) \cite{jhtdb}. It consists of a DNS simulation using a pseudo-spectral method with 1024$^3$ nodes in a periodic domain of size $(0,2\pi)^3$ over $T=10$~s, with 5028 timesteps. Energy is injected to maintain a constant total energy in modes such that their wave number magnitude is less than or equal to 2. We use a smaller derived dataset, taking 500 snapshots at a smaller resolution of $128^3$ using spectral downsampling. 
Finally, we shuffle all snapshots and use 400 and 100 samples for training and testing, respectively. The divergence of the flow field $\nabla \cdot \mathbf{u}$ is used as the metric for physics consistency.
Figures \ref{fig:example_2D_data} and \ref{fig:v_original} show samples of our training data.

\subsection{Physcial-consistency}
\label{sec:train_meth}
We first target an unconditioned generation task to examine the ability of different methods to generate physically constrained samples. We consider standard diffusion models, Physics-Informed Diffusion Models~\cite{bastek2024physics} with dynamic rescale factor ($\text{PIDM}_\text{dyn}$), ConFIG~\cite{liu2024config}, and our $\text{ConFIG}_\text{u}$. Some comments about training hyperparameters and the network's architecture can be found in Appendix \ref{sec:hyper-params}. 

To calculate the physical residual, the time and spatial derivatives are evaluated in a discrete manner. Ideally, the physical residual of the training data, $\Vert\mathcal{R}(\omega^*)\Vert_2$, should be 0. Still, we usually encounter a non-zero physical residual due to the discretisation errors. As no method can be expected to completely resolve this discretization error, it obscures the capabilities of the methods to improve the diffusion training.
Hence, we employ a metric that compares the residual of the generated samples $\omega'$ to the average of the test dataset samples:
\begin{equation}
    \boldsymbol{L_\text{\textbf{res}}} = | \Vert \mathcal{R}(\omega') \Vert_2 - \mathcal{R}_{\text{avg}}^*|
\end{equation}
where $\mathcal{R}_{\text{avg}}^* := \frac{1}{N}\sum_{i}^N \Vert\mathcal{R}(\omega^*)\Vert_2$ is the average physical residual of the test dataset.

\begin{ruledtabular}
\begin{table*}[ht!]
    \centering
    \begin{tabular}{c c c c c}
        \textbf{Test case} & \textbf{Training method} & \textbf{$\boldsymbol{L_\text{\textbf{res}}}$} & \textbf{$\Vert \mu' - \mu^* \Vert_2$} \\ 
        
        \hline
        
        \multirow{4}{*}{2D Kolmogorov flow} & Std. Diffusion & 1.27 $\pm$ 0.12 & 0.455 $\pm$ 0.104 \\ 
        
        & $\text{PIDM}_{\text{dyn}}$~\cite{bastek2024physics} & \textbf{0.99 $\pm$ 0.21} & \textbf{0.316 $\pm$ 0.005} \\
        
        & ConFIG~\cite{liu2024config} & 1.24 $\pm$ 0.32 & 0.340 $\pm$ 0.009 \\
        
        & $\text{ConFIG}_\text{u}$ & \best{0.68 $\pm$ 0.15} & \best{0.309 $\pm$ 0.008} \\ 
\hline
        \multirow{4}{*}{3D isotropic turbulence}& Std. Diffusion & 1.03 $\pm$ 0.07 & 0.690 $\pm$ 0.037 \\ 
        
        &$\text{PIDM}_{\text{dyn}}$~\cite{bastek2024physics} & 1.08 $\pm$ 0.07 & 0.696 $\pm$ 0.042 \\

        & ConFIG~\cite{liu2024config} & \textbf{1.01 $\pm$ 0.05} & \textbf{0.667 $\pm$ 0.067} \\
        
        & $\text{ConFIG}_\text{u}$ & \best{0.80 $\pm$ 0.05} & \best{0.663 $\pm$ 0.046} \\ 
        
    \end{tabular}
    \caption{Evaluation of the training methods under consideration. $\mu'$ and $\mu^*$ represent the pixel-wise mean of the generated and the test samples, respectively.}
    \label{tab:gen_metrics}
\end{table*}
\end{ruledtabular}

The evaluations of the different training methods can be found in Table~\ref{tab:gen_metrics}, where the $\text{ConFIG}_\text{u}$ method reaches the best results. Specifically for the physical residual metric, the ConFIG$_\text{u}$ achieves an improvement of 50\% and 22\% compared to standard diffusion models in the 2D and 3D cases, respectively. It also outperforms the regular ConFIG method by similar margins. Hence, we will focus on employing the ConFIG$_\text{u}$ method in the following experiments. 

\subsection{Reconstruction tasks}
\label{sec:sample_meth}
\begin{figure}[ht!]
    \centering
    \includegraphics[width=\linewidth]{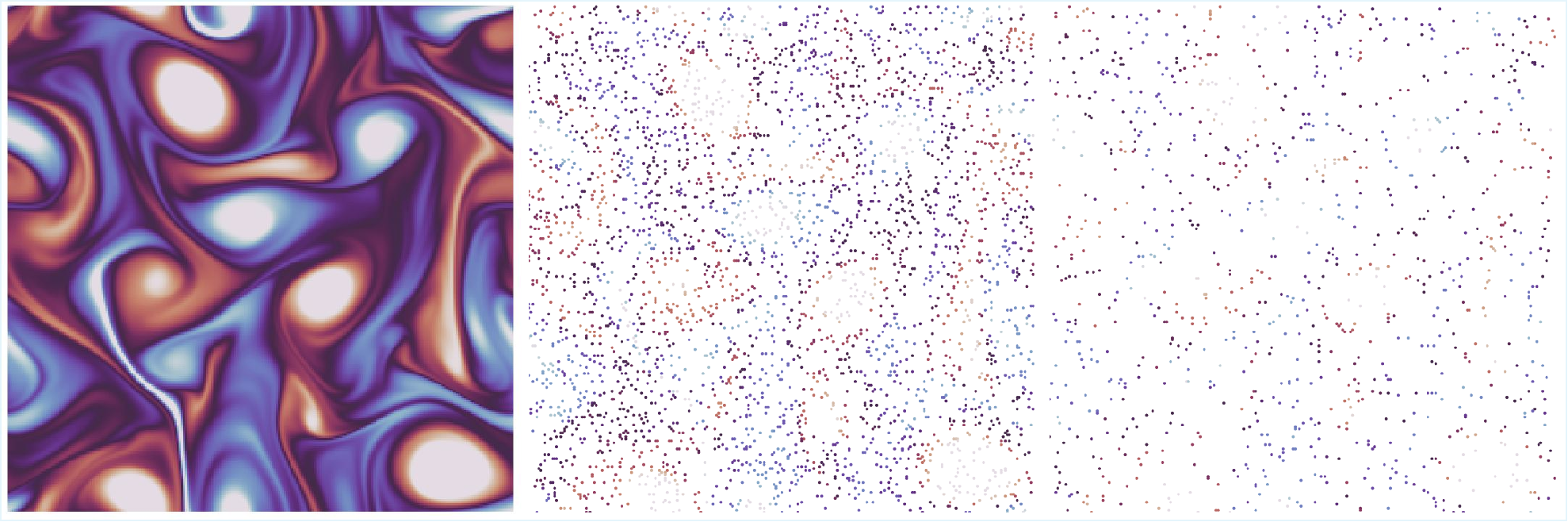}
    \caption{From left to right: a sample of the dataset, and samples containing 5\% and 1.5625\% of the data, respectively.}
    \label{fig:sparse_example}
\end{figure}

Next, we test the proposed masked diffusion method for reconstructing high-fidelity flow fields from sparse data points. Following the configuration from the previous study by Shu et al. \cite{shu2023physics}, we target reconstruction tasks with sparse observations of 5\% and 1.5625\% of the high fidelity data. An example of the sparse input can be seen in Figure \ref{fig:sparse_example}.

The reconstructed samples should meet three key properties: accurate point-wise values, physical coherence, and a similar fluid structure to the ground truth. 

The first property is typically measured using the Root Mean Square Error (RMSE) between the predicted flow field and the ground truth. To satisfy physical coherence, the reconstructed field should follow the physical residual of the data, which can be evaluated using the relative residual loss $L_\text{\textbf{res}}$ introduced in Section~\ref{sec:train_meth}. 

Evaluating the structure of a fluid flow improves our understanding of how closely the predicted sample aligns with the target high-fidelity field. As no ground truth is available in this probabilistic setting, we use a learned, perceptual similarity metric (LSiM) \cite{kohl2020learning} to measure the similarity between our reconstruction and the ground truth. This metric allows us to assess whether the structural characteristics of the generated sample align with those of the target field. 

We also use the error in the energy spectrum to analyse errors in the frequency domain. For the 3D case, we compute the L1 integral norm in logarithmic scale of the difference between the ground truth and the reconstructed spectra, given by:
\begin{equation}
    \int_{k_{min}}^{k_{max}} \left| \log E(k) - \log E_{GT}(k) \right| d(\log k)
\end{equation}
For the 2D scenario, we compute this error using the enstrophy spectrum instead of the energy spectrum, as used in other works \cite{chen2024probabilistic}. This is because enstrophy cascades to small scales in 2D, providing a more sensitive and physically meaningful diagnostic of whether the simulation captures the correct multiscale turbulent dynamics.

\begin{figure*} [ht!]
    \centering
  \begin{subfigure}[b]{0.49\linewidth}
    \includegraphics[width=0.8\linewidth]{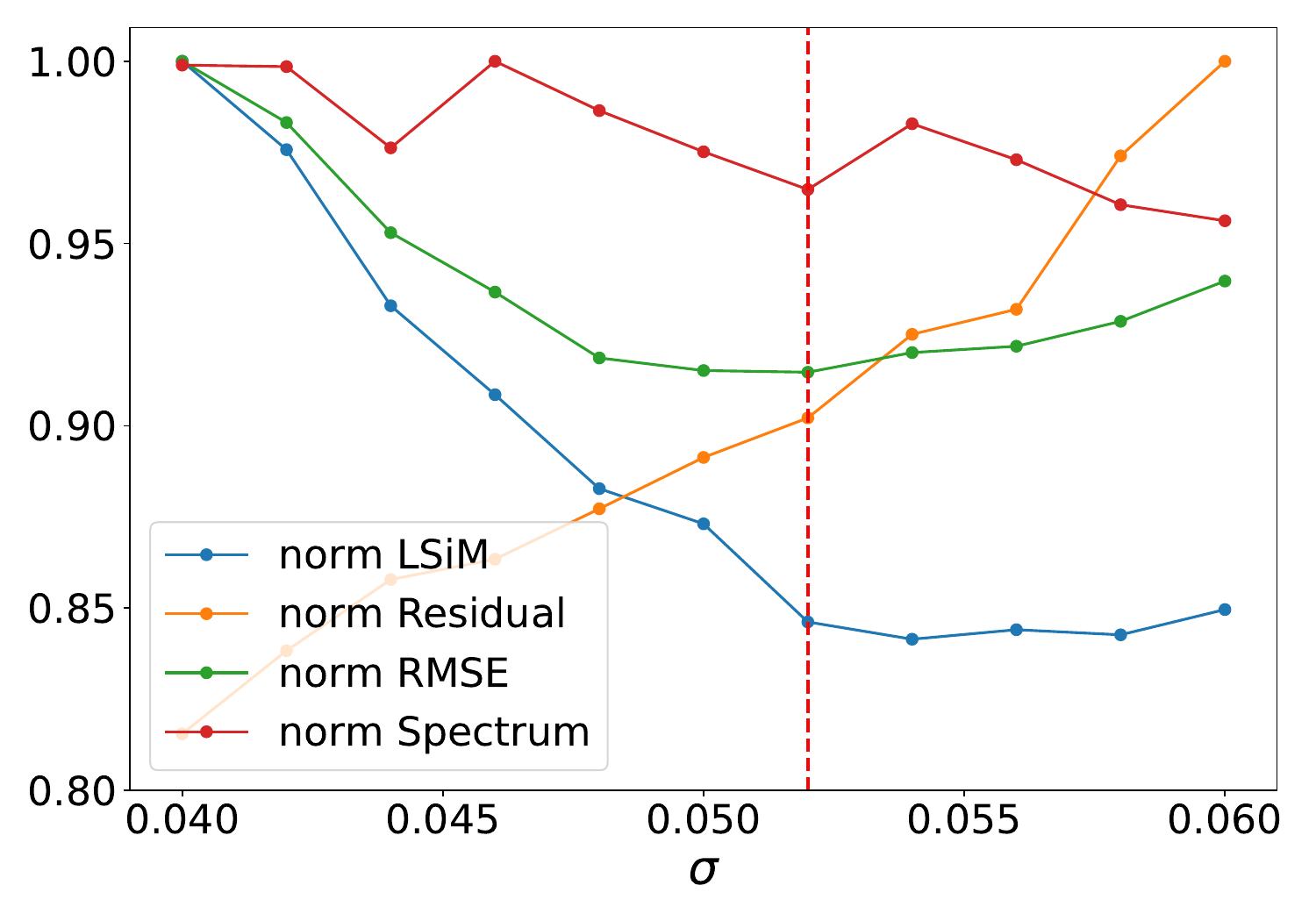}
    \caption{1.5625\% task, sigma fixed at $\sigma = 0.052$.}
  \label{fig:sig:a}
  \end{subfigure}%
  \begin{subfigure}[b]{0.49\linewidth}
    \includegraphics[width=0.8\linewidth]{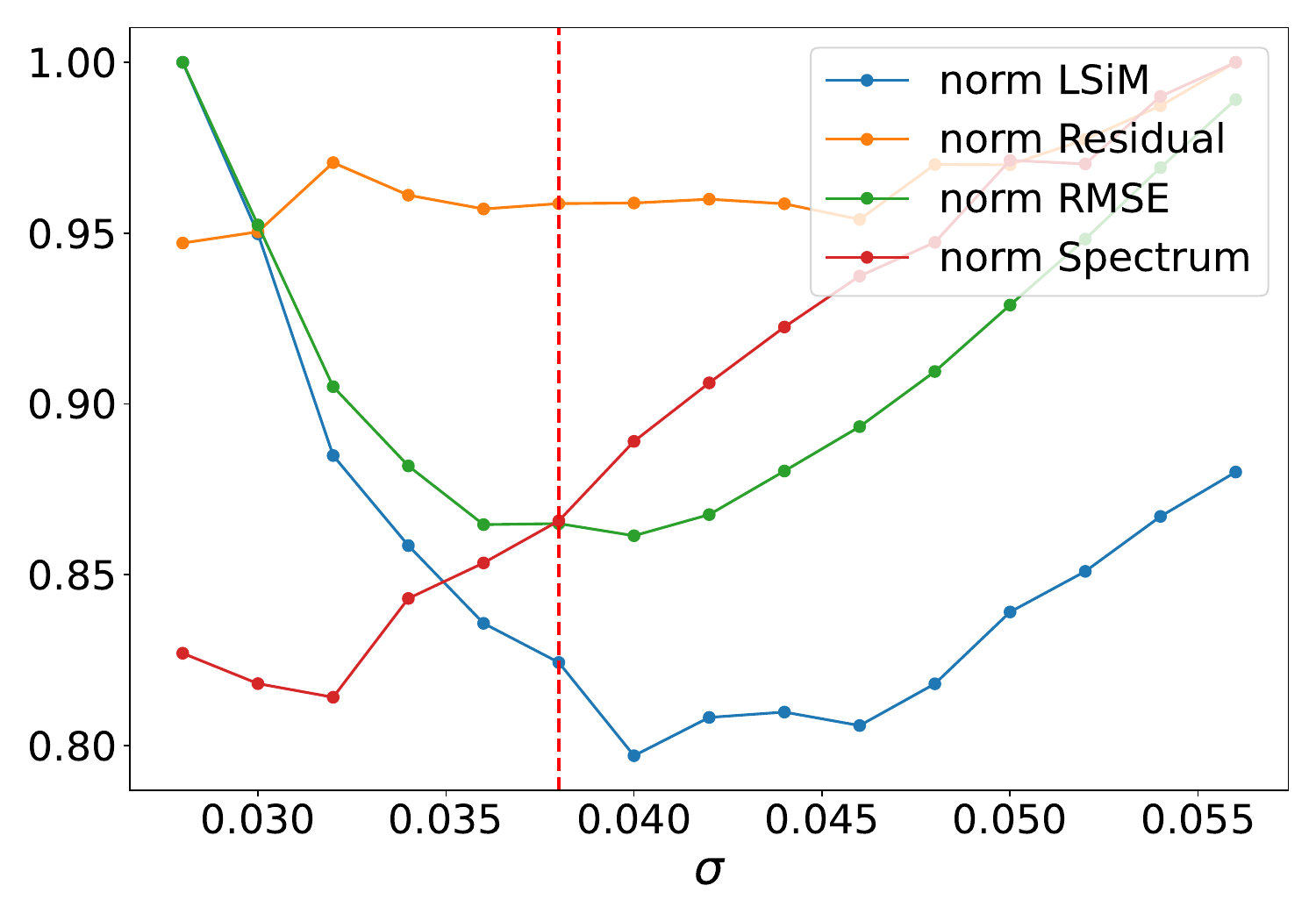}
    \caption{5\% task, sigma fixed at $\sigma = 0.038$.}
  \label{fig:sig:b}
  \end{subfigure}
  \caption{Sigma value exploration for the 2D Kolmogorov case. The red vertical line indicates the chosen value for the parameter. All metrics are normalised.}
  \label{fig:sigma}
\end{figure*}

\begin{figure*}[ht!]
\centering
\begin{subfigure}{0.49\linewidth}
    \includegraphics[width=0.8\linewidth]{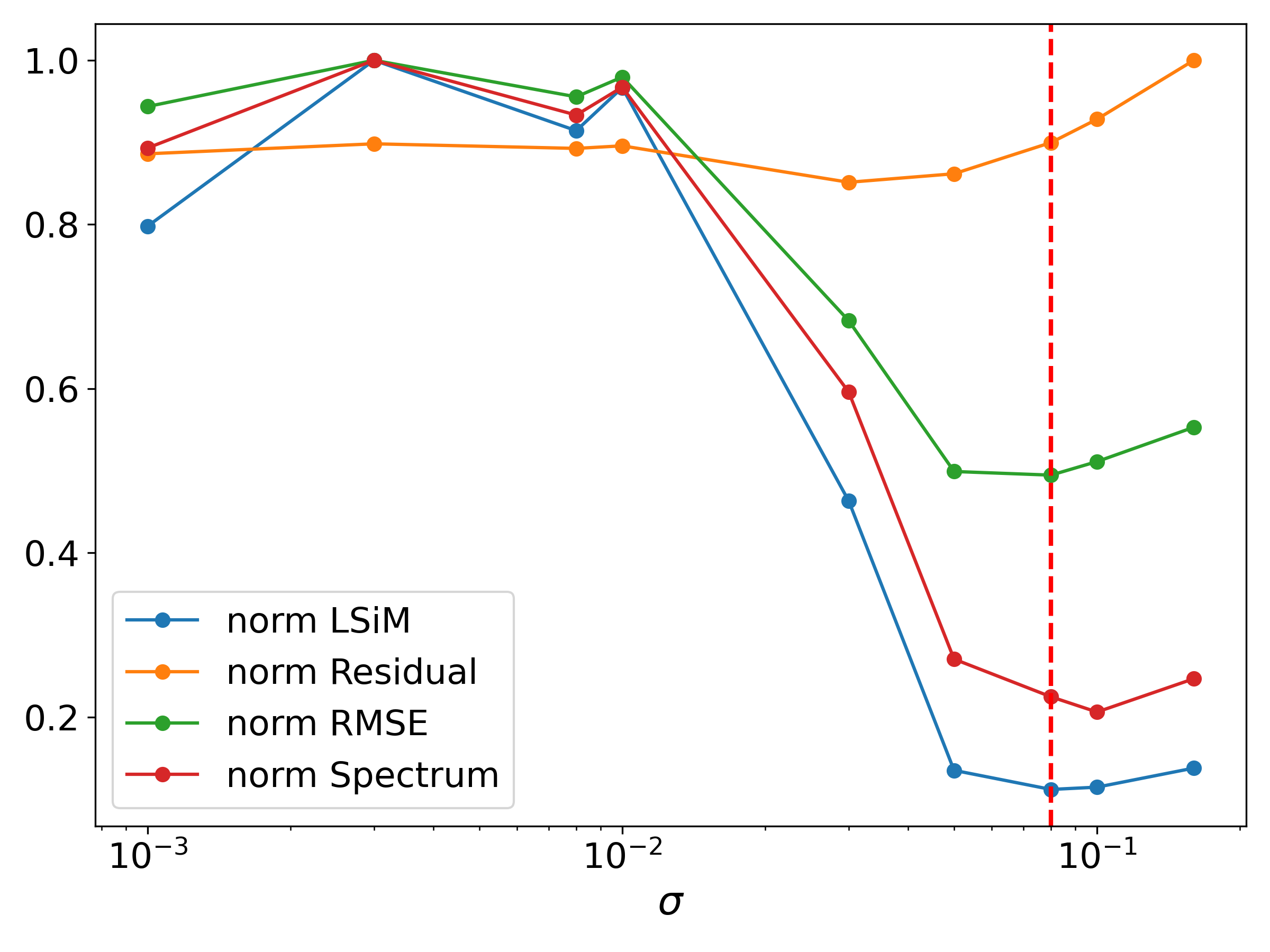}
    \caption{1.5625\% task, sigma fixed at $\sigma=0.08$}
    \label{fig:sigma2}
\end{subfigure}
\begin{subfigure}{0.49\linewidth}
    \includegraphics[width=0.8\linewidth]{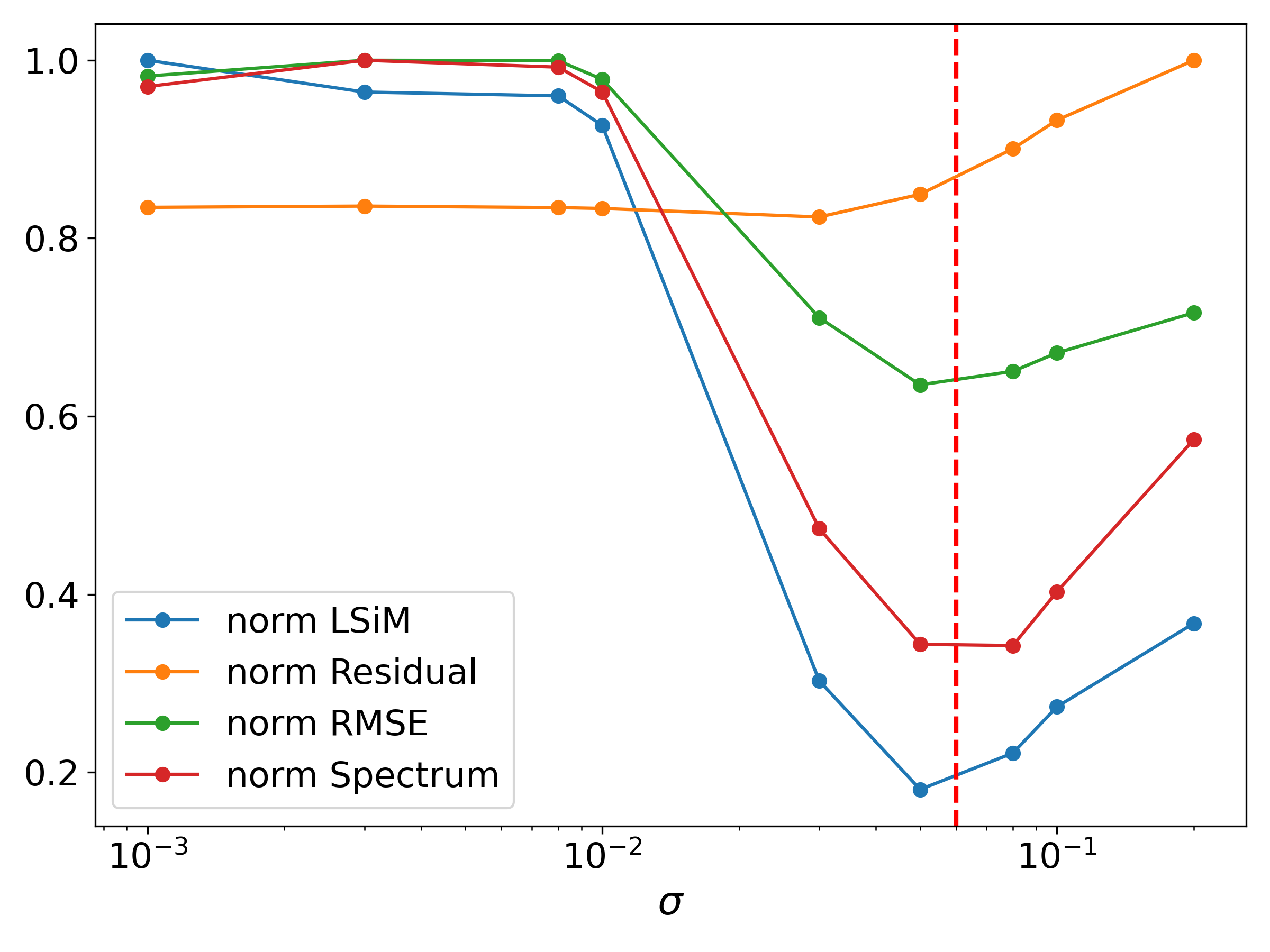}
        \caption{5\% task, sigma fixed at $\sigma=0.06$}
        \label{fig:sigma1}
\end{subfigure}
\caption{Sigma value exploration for the 3D isotropic turbulence case. All metrics are normalised.}
\label{fig:sigma_3d}
\end{figure*}

The smoothing coefficient of the mask $\sigma$ is a hyperparameter in our method. It determines the width of the Gaussian functions of the mask, centred at every value of the sparse data. We conducted a parameter exploration, shown in Figure~\ref{fig:sigma} and Figure~\ref{fig:sigma_3d}, which allowed us to obtain the values of $\sigma = 0.038$ and $\sigma = 0.052$ for the 5\% and 1.5625\% tasks in 2D, respectively, and $\sigma = 0.06$ and $\sigma = 0.08$ in 3D. Meanwhile, we additionally use a scheduler to control the strength of the mask in our sampling approach, as discussed in Section~\ref{sec:meth_sampling}. Experimentally, we observed that the scheduler $\gamma_t = \left(t/T\right)^3$ gave satisfactory results.

Besides comparing our method with that of Shu et al. \cite{shu2023physics}, we also compare our reconstruction result with a supervised baseline where the network directly predicts high-fidelity data from an interpolated sparse input. For a fair comparison, we use the same backbone architectures as the ones for the diffusion models. Moreover, we need to highlight that the baseline and other diffusion models for sparse data reconstruction require specific training for each new scenario (different \% of data, for instance) to maximize their performance. In contrast, the masked diffusion method is general and only needs to be trained once, as an unconditional generator, using only the high-resolution training samples.
Table~\ref{tab:task5} shows the results of the 2D 5\% reconstruction task, where we can observe how our masked diffusion method outperforms previous methods on the RMSE, residual, structure similarity, and energy spectrum errors.
We also provide the \textit{P-RMSE} values, which measure the point-wise accuracy only at the sparse input data locations. These P-RMSE measurements highlight the improvements that masked diffusion provides at the locations constrained by input data.

Similar results can be observed when evaluating the 1.5625\% task, which can be seen in Table~\ref{tab:task1}. Our method consistently obtains the lowest errors in both tasks. 

\begin{table*}[ht!]
\begin{ruledtabular}
    \centering
    \begin{tabular}{ l c c c c c }
    \textbf{Models} & \textbf{RMSE} & \textbf{P-RMSE} & \textbf{$\boldsymbol{L_\text{\textbf{res}}}$} & \textbf{LSiM} & \textbf{E.spectrum}\\
    \hline
    Supervised baseline & 0.79 $\pm$ 0.09 & 0.79 $\pm$ 0.01 & \textbf{0.16 $\pm$ 0.07} & 0.29 $\pm$ 0.02 &  \textbf{0.46 $\pm$ 0.12} \\
    Shu et al. \cite{shu2023physics} & \textbf{0.18 $\pm$ 0.03} & \textbf{0.14 $\pm$ 0.02} & 0.84 $\pm$ 0.07 & \textbf{0.10 $\pm$ 0.02} & 0.54 $\pm$ 0.08 \\
    Masked diffusion (ours) & \best{0.13 $\pm$ 0.02} & \best{0.10 $\pm$ 0.01} & \best{0.02 $\pm$ 0.03} & \best{0.05 $\pm$ 0.01} & \best{0.26 $\pm$ 0.05} \\
\end{tabular}
\caption{Evaluation of the reconstruction task from 5\% of the data for 2D Kolmogorov flow.}
\label{tab:task5}    
\end{ruledtabular}
\end{table*}

\begin{table*}[ht!]
\begin{ruledtabular}
    \centering
    \begin{tabular}{ l c c c c c }
    \textbf{Models} & \textbf{RMSE} & \textbf{P-RMSE} & \textbf{$\boldsymbol{L_\text{\textbf{res}}}$} & \textbf{LSiM} & \textbf{E.spectrum}\\
    \hline
    Supervised baseline & 1.06 $\pm$ 0.16 & 0.81 $\pm$ 0.02 & \textbf{0.21 $\pm$ 0.14} & 0.42 $\pm$ 0.06 & 0.68 $\pm$ 0.23 \\
    Shu et al. \cite{shu2023physics} & \textbf{0.37 $\pm$ 0.06} & \textbf{0.22 $\pm$ 0.03} & 0.22 $\pm$ 0.11 & \textbf{0.17 $\pm$ 0.03} & \textbf{0.52 $\pm$ 0.12} \\
    Masked diffusion (ours) & \best{0.31 $\pm$ 0.06} & \best{0.15 $\pm$ 0.02} & \best{0.15 $\pm$ 0.06} & \best{0.11 $\pm$ 0.02} & \best{0.33 $\pm$ 0.07} \\
    \end{tabular}
    \caption{Evaluation of the reconstruction task from 1.5625\% of the data for 2D Kolmogorov flow.}
    \label{tab:task1}
\end{ruledtabular}
\end{table*}

\begin{table*}[ht!]
\begin{ruledtabular}

    \centering
    \begin{tabular}{ l c c c c c}
    \textbf{Models} & \textbf{RMSE} & \textbf{P-RMSE} & \textbf{$\boldsymbol{L_\text{\textbf{res}}}$} & \textbf{LSiM} & \textbf{E.spectrum}\\
    \hline
    Supervised baseline & \best{0.13 $\pm$ 0.01} & 0.22 $\pm$ 0.01 & \best{0.32 $\pm$ 0.02} & \best{0.05 $\pm$ 0.01} & 0.64 $\pm$ 0.02 \\
    Shu et al. \cite{shu2023physics} & 0.45 $\pm$ 0.03 & 0.68 $\pm$ 0.03 & 0.80 $\pm$ 0.06 & 0.64 $\pm$ 0.15 & 0.38 $\pm$ 0.07 \\
    Masked diffusion (ours) & 0.17 $\pm$ 0.01 & \best{0.11 $\pm$ 0.01} & 0.61 $\pm$ 0.03 & \textbf{0.06 $\pm$ 0.01} & \best{0.05 $\pm$ 0.01} \\
    \end{tabular}
    \caption{Evaluation of the reconstruction task from 5\% of the data for the 3D turbulent case.}
    \label{tab:task7}    
\end{ruledtabular}
\end{table*}

\begin{table*}[ht!]
\begin{ruledtabular}
    \centering
    \begin{tabular}{ l c c c c c}
    \textbf{Models} & \textbf{RMSE} & \textbf{P-RMSE} & \textbf{$\boldsymbol{L_\text{\textbf{res}}}$} & \textbf{LSiM} & \textbf{E.spectrum}\\
    \hline
    Supervised baseline & \best{0.16 $\pm$ 0.01} & 0.27 $\pm$ 0.01 & \best{0.36 $\pm$ 0.01} & \best{0.08 $\pm$ 0.01} & 0.71 $\pm$ 0.02 \\
    Shu et al. \cite{shu2023physics} & 0.44 $\pm$ 0.02 & 0.69 $\pm$ 0.04 & 0.79 $\pm$ 0.05 & 0.62 $\pm$ 0.11 & 0.36 $\pm$ 0.04 \\
    Masked diffusion (ours) & 0.20 $\pm$ 0.01 & \best{0.18 $\pm$ 0.01} &  0.66 $\pm$ 0.03 & \textbf{0.09 $\pm$ 0.01} & \best{0.09 $\pm$ 0.01} \\
    \end{tabular}
    \caption{Evaluation of the reconstruction task from 1.5625\% of the data for the 3D turbulent case.}
    \label{tab:task8}    
\end{ruledtabular}
\end{table*}

\begin{figure*}[ht!]
\centering
\begin{subfigure}[b]{0.48\linewidth}
    \includegraphics[width=0.9\linewidth]{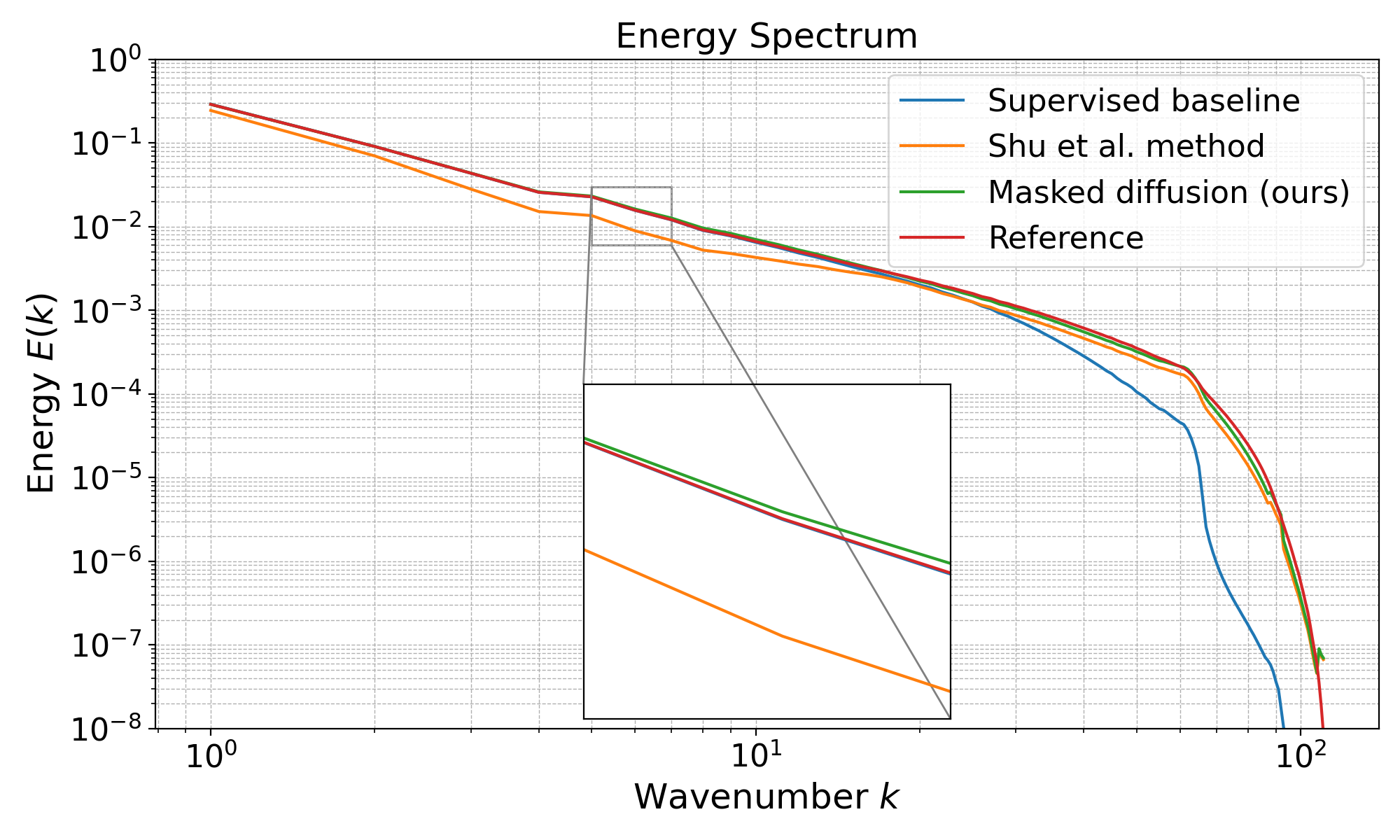}
        \caption{5\% sparse data reconstruction.}
        \label{fig:q_5}
\end{subfigure}
\begin{subfigure}[b]{0.48\linewidth}
    \includegraphics[width=0.9\linewidth]{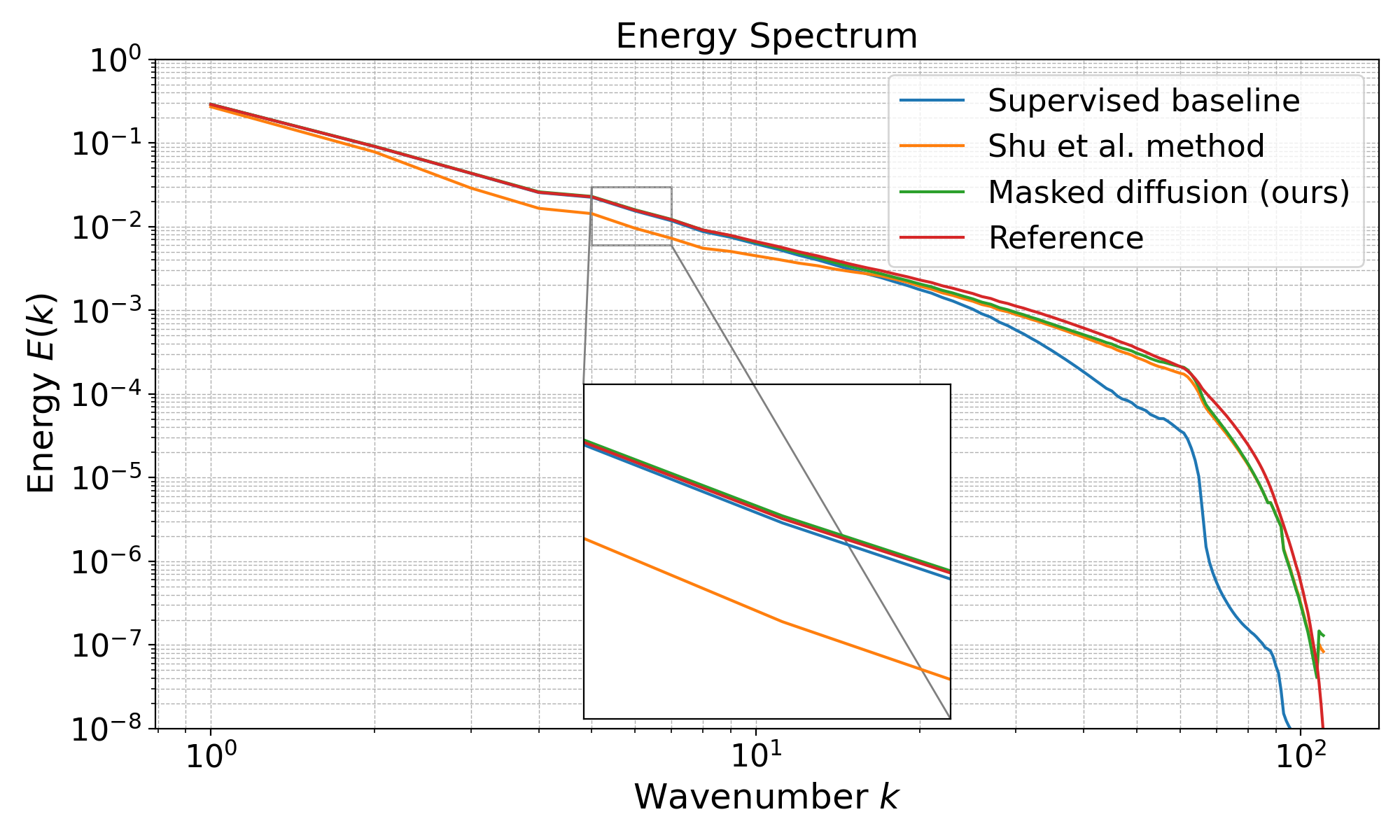}
        \caption{1.5625\% sparse data reconstruction.}
        \label{fig:q_4}
\end{subfigure}
\caption{Energy spectrum comparison of different reconstruction methods for our two 3D scenarios.}
\label{fig:freq_dist}
\end{figure*}

In Tables~\ref{tab:task7} and \ref{tab:task8} we show the results for the 3D case of the same two reconstruction tasks. Our masked diffusion method outperforms Shu et al.’s method across all evaluation metrics. However, the supervised baseline achieves lower errors in RMSE, physical residual, and LSiM. These metrics are primarily influenced by the accuracy of the low-frequency components, which the supervised baseline captures more effectively. As shown in Figure \ref{fig:freq_dist}, only the diffusion methods are capable of accurately reconstructing the high-frequency components. In contrast, the supervised baseline produces noticeably blurrier results despite achieving a lower RMSE. The spectrum in Figure \ref{fig:freq_dist} shows that the high wavenumber content of the solutions inferred by the supervised baseline clearly lacks energy. Another drawback of the supervised baseline (and other reconstruction methods like diffusion with conditioning on the low-resolution sample) is that it needs to be re-trained if the amount of available observations changes, while the method by Shu et al.~\cite{shu2023physics} and our masked diffusion only need to be trained once using only the high-resolution samples.
Besides, as expected, our method achieves the lowest P-RMSE at the sparse data locations among the tested approaches.
The masked diffusion method shown in these plots uses nearest interpolation to start the guidance in masked diffusion (our default case). 
However, one could use other, more advanced interpolation techniques for this purpose. We analyze an alternative in Appendix~\ref{app:interpolation}.

\begin{figure*}[ht!]
    \centering
    \includegraphics[width=0.7\textwidth]{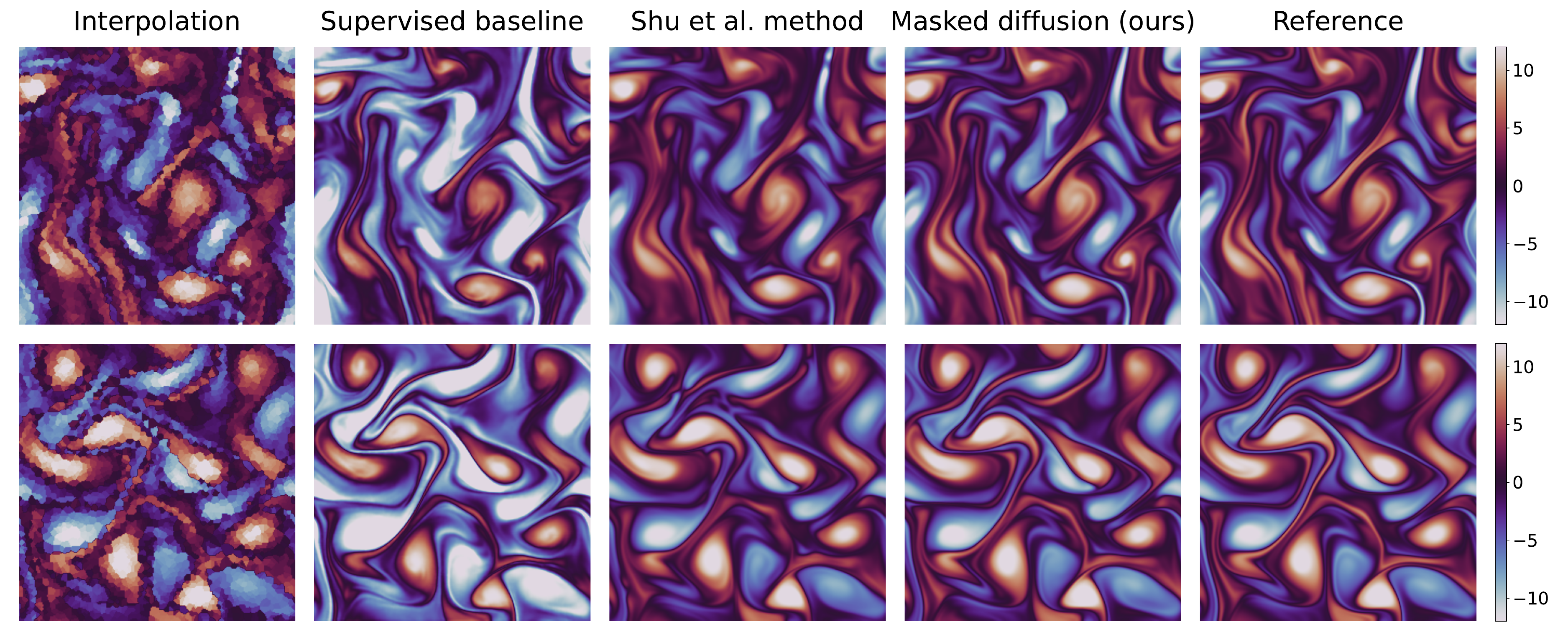}
    \caption{Example 2D Kolmogorov flow 5\% data reconstruction. }
    \label{fig:example5}
\end{figure*}

\begin{figure*}[ht!]
    \centering
    \includegraphics[width=0.7\textwidth]{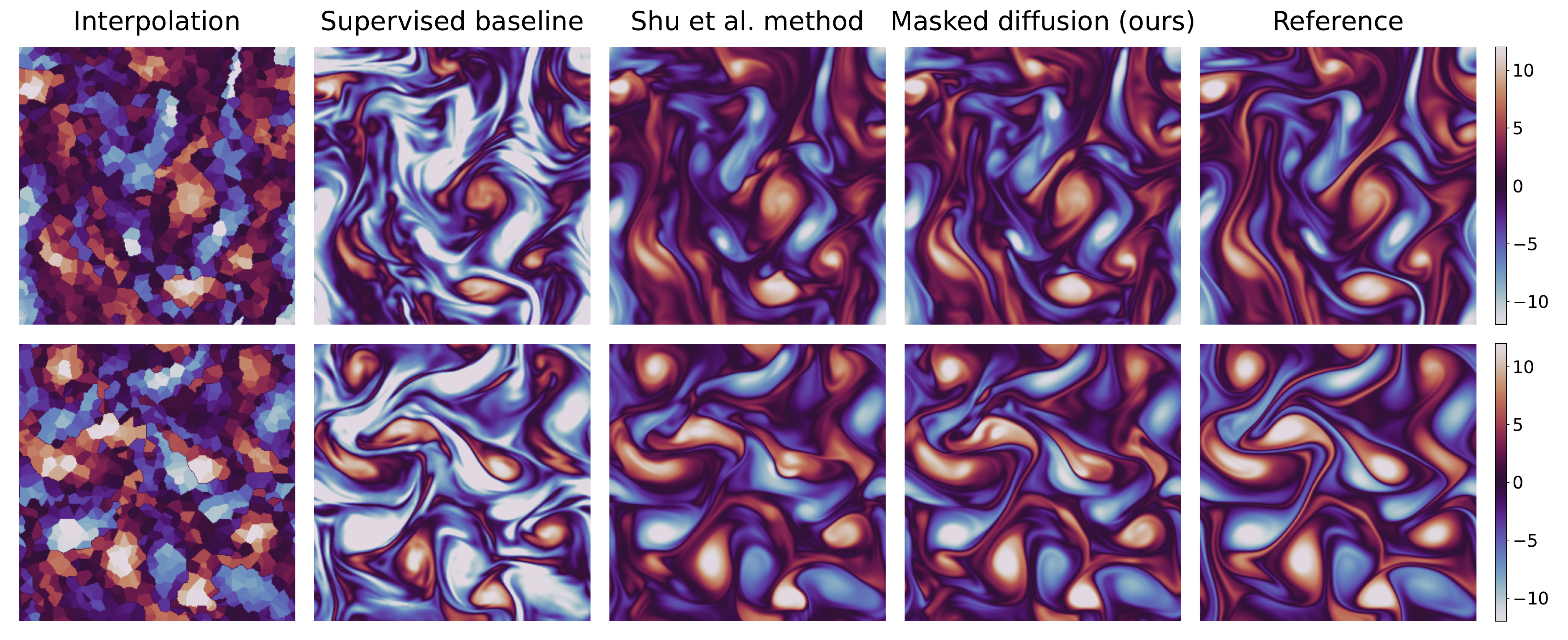}
    \caption{Example 2D Kolmogorov flow 1.5625\% data reconstruction.}
    \label{fig:example1}
\end{figure*}

\begin{figure*}[ht!]
    \centering
    \includegraphics[width=0.75\linewidth]{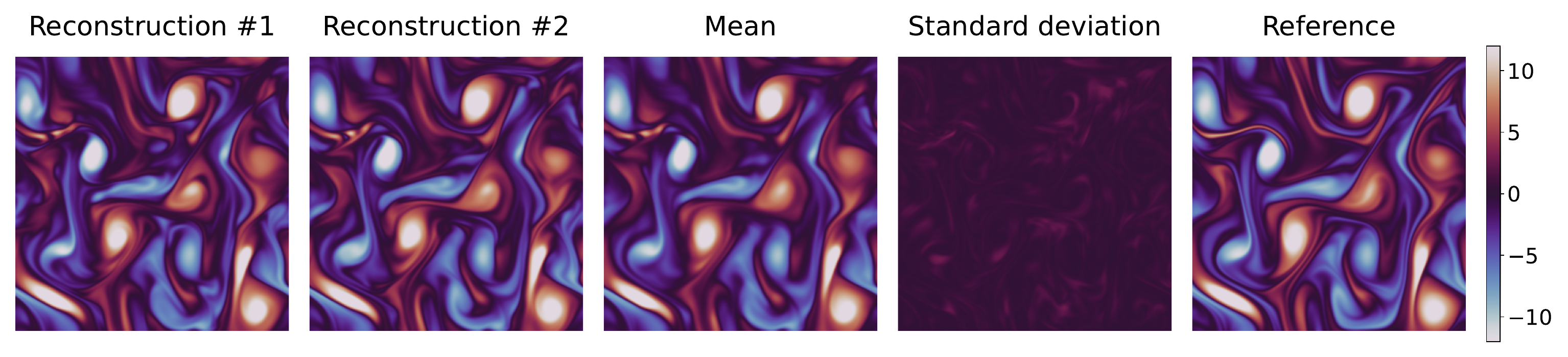}
    \caption{Example of variability when reconstructing samples with 1.5625\% of data. Mean and standard deviation were computed from 10 reconstructions. }
    \label{fig:variety_reconstr}
\end{figure*}

\begin{figure*}[ht!]
  \centering
  \includegraphics[width=0.65\textwidth]{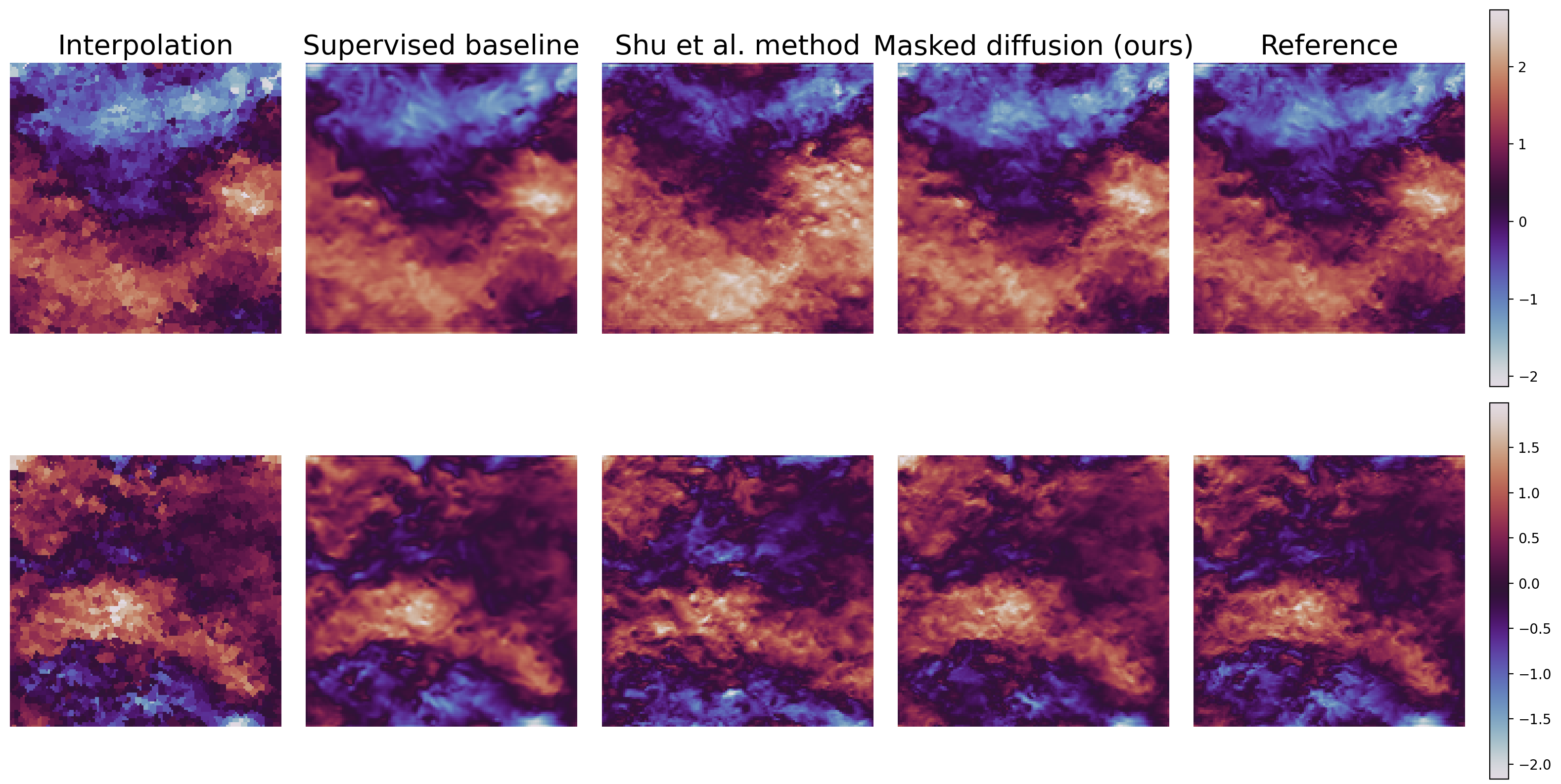}
  \caption{Example 3D isotropic turbulence 5\% sparse data reconstruction (2D slices).}
  \label{fig:reg5}
\end{figure*}

\begin{figure*}[ht!]
  \centering
  \includegraphics[width=0.65\textwidth]{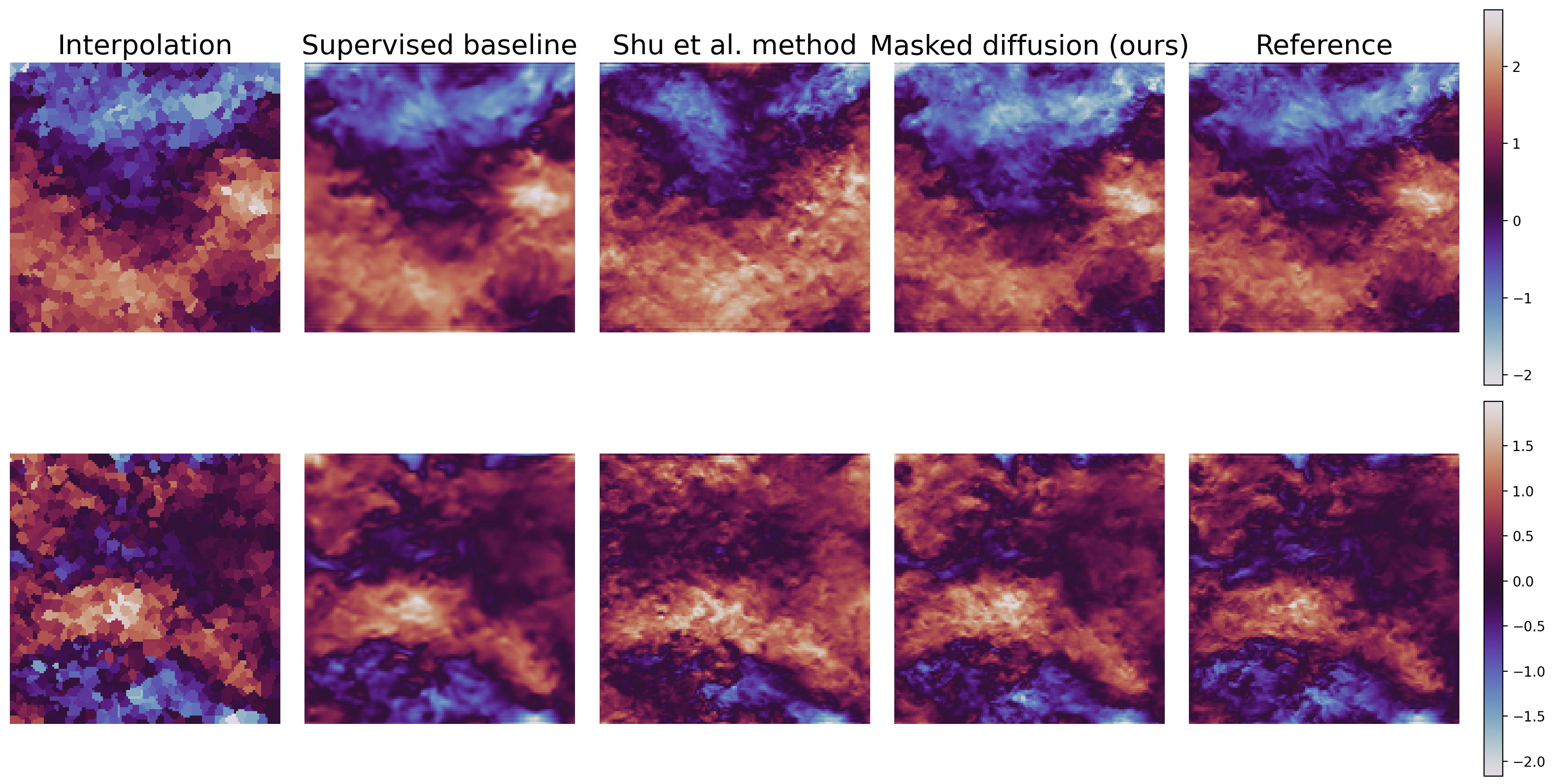}
  \caption{Example 3D isotropic turbulence 1.5625\% sparse data reconstruction (2D slices).}
  \label{fig:reg4}
\end{figure*}

Figures~\ref{fig:example5}-\ref{fig:reg4} show samples of the reconstruction in 2D and 3D cases. It can be seen how masked diffusion better approximates the complicated structures of the turbulent flow fields, producing sharper images and yielding a closer approximation of the flow structures and statistics. In addition, since the diffusion model is inherently generative, it can produce diverse realizations for the same sparse input, as we show in Figure \ref{fig:variety_reconstr}. The low deviation between samples highlights the high confidence of the model in the proposed reconstruction.

The advantage of the supervised baseline stems from its training strategy: the network learns to interpolate missing data points directly from sparse observations. This interpolation performs well for large-scale structures but fails to reliably capture small-scale features. However, reconstructing high-frequency components is particularly important in turbulent simulations, where dynamics span a wide range of physical scales. For this reason, we consider the spectrum error as the central performance metric.

\begin{figure}[ht!]
\centering
\includegraphics[width=0.8\linewidth]{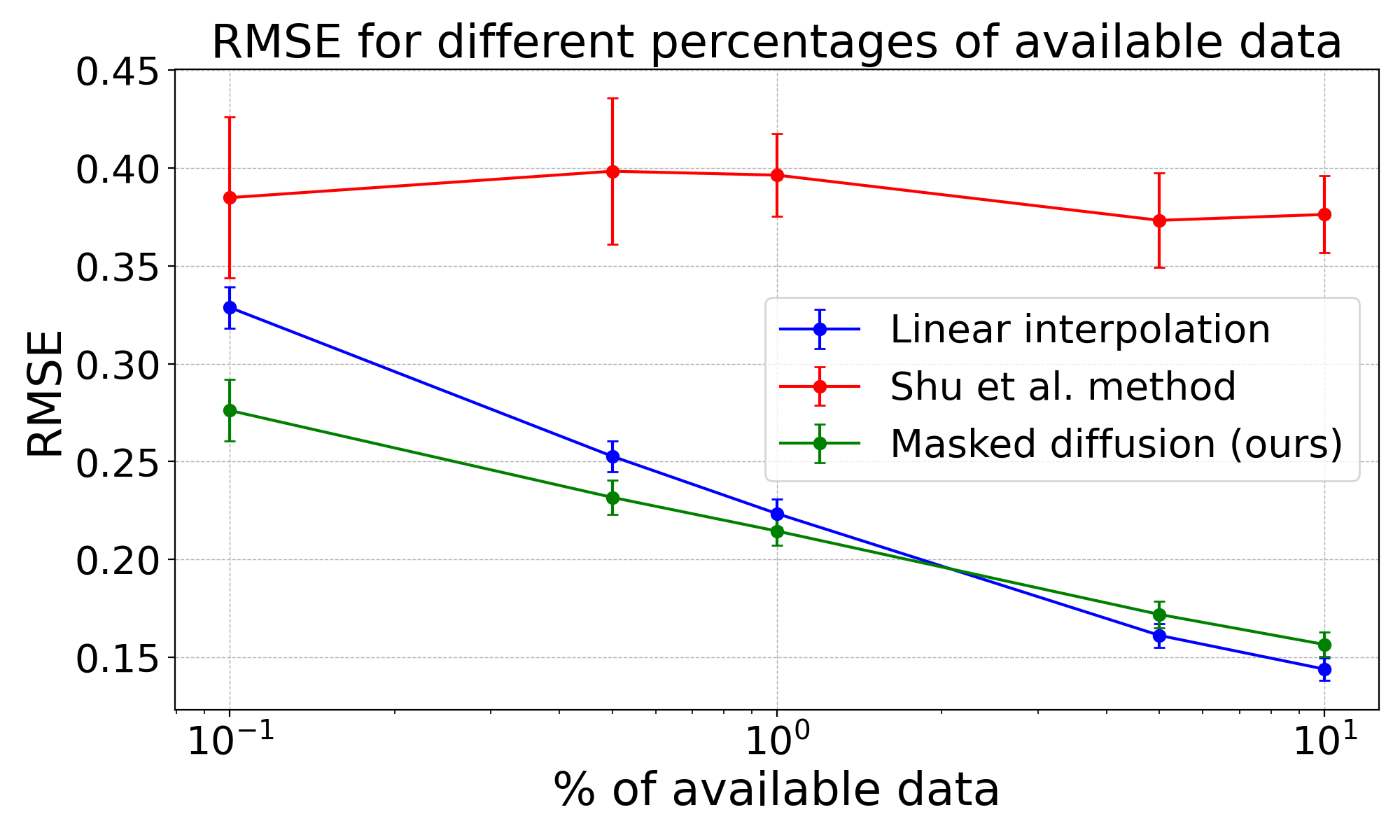}
    \caption{RMSE of the 3D isotropic flow reconstruction with varying percentages of sparse data.}
\label{fig:rmse_distance_perc}
\end{figure}

\begin{figure}[ht!]
\centering
\includegraphics[width=0.8\linewidth]{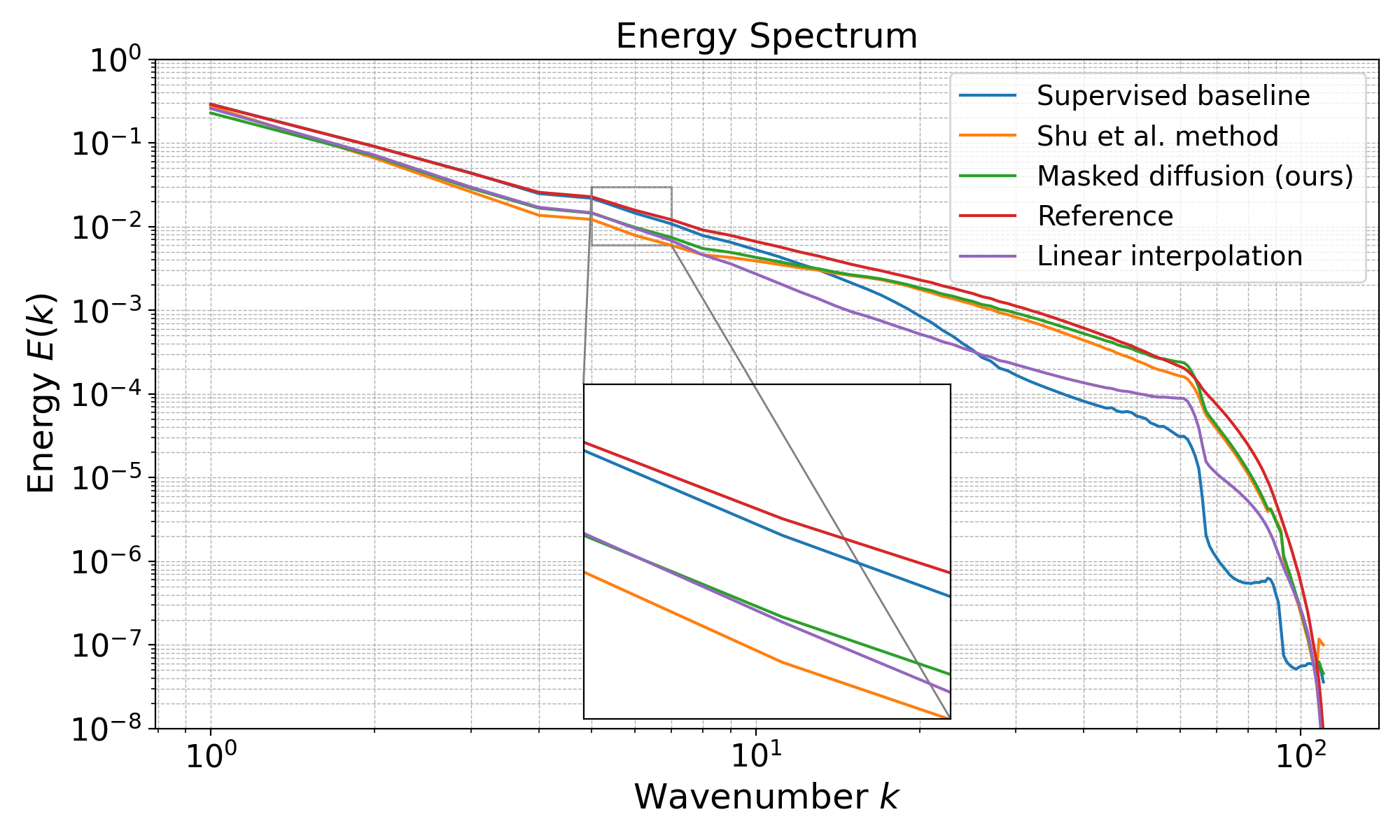}
    \caption{Energy spectrum comparison for a 0.1\% 3D sparse data reconstruction.}
\label{fig:rmse_spectrum_point1}
\end{figure}

As is evident from the spectrum errors in Tables~\ref{tab:task7} and \ref{tab:task8}, only the two considered diffusion models are capable of capturing small-scale features. E.g., our method features more than $6.4\times$ lower error than the supervised baseline. However, the two diffusion approaches behave very differently when considering different amounts of available data. In Figure~\ref{fig:rmse_distance_perc}, we plot the RMSE over the 3D reconstruction of 50 samples with Shu et al.'s method and our masked diffusion method (with nearest interpolation) together with direct linear interpolation as a baseline. Naturally, a reduced performance can be expected for very low amounts of input data. However, the two learned methods vary in their ability to handle the low-data regime.
We observe that Shu et al.'s method remains mostly constant, with a consistently higher RMSE than our method. We see that if the percentage of available data is too low, our method achieves even lower RMSE errors than linear interpolation, in addition to producing high-frequency content. As a particularly challenging test case, table \ref{tab:task_point1}  shows the same metrics for a case with only 0.1\% of sparse input data.
Masked diffusion is the best at reconstructing the high-frequency components of the spectrum, as shown in Figure \ref{fig:rmse_spectrum_point1} for the challenging 0.1\% data case. We include visualizations of several example reconstructions in Appendix \ref{app:extreme}.
Thus, the proposed method performs very well across the different regimes of data availability, for high percentage cases where linear interpolation and the supervised baseline still produce inaccurate high-wavenumber components, as well as highly sparse scenarios.

\begin{table*}[ht!]
\begin{ruledtabular}
    \vspace{1cm} 

    \centering
    \begin{tabular}{ l c c c c c}
    \textbf{Models} & \textbf{RMSE} & \textbf{P-RMSE} & \textbf{$\boldsymbol{L_\text{\textbf{res}}}$} & \textbf{LSiM} & \textbf{E.spectrum}\\
    \hline
    Linear interpolation & 0.33 $\pm$ 0.01 & \best{0.00 $\pm$ 0.01} & 1.75 $\pm$ 0.06 & 0.42 $\pm$ 0.04 & 0.75 $\pm$ 0.02 \\
    Supervised baseline & \best{0.21 $\pm$ 0.01} & 0.36 $\pm$ 0.01 & \best{0.59 $\pm$ 0.01} & \best{0.17 $\pm$ 0.02} & 0.74 $\pm$ 0.02 \\
    Shu et al. \cite{shu2023physics} & 0.39 $\pm$ 0.04 & 0.69 $\pm$ 0.06 & \textbf{0.67 $\pm$ 0.03} & 0.46 $\pm$ 0.08 & \textbf{0.32 $\pm$ 0.02} \\
    Masked diffusion (ours) & \textbf{0.27 $\pm$ 0.02} & \textbf{0.18 $\pm$ 0.01} & 0.74 $\pm$ 0.03 & \textbf{0.24 $\pm$ 0.08} & \best{0.23 $\pm$ 0.02} \\
    \end{tabular}
    \caption{Evaluation of the reconstruction task from 0.1\% of the data for the 3D turbulent case.}
    \label{tab:task_point1}    
\end{ruledtabular}
\end{table*}

\begin{figure*}[ht!]
\centering
\begin{subfigure}[b]{0.49\linewidth}
    \includegraphics[width=0.85\linewidth]{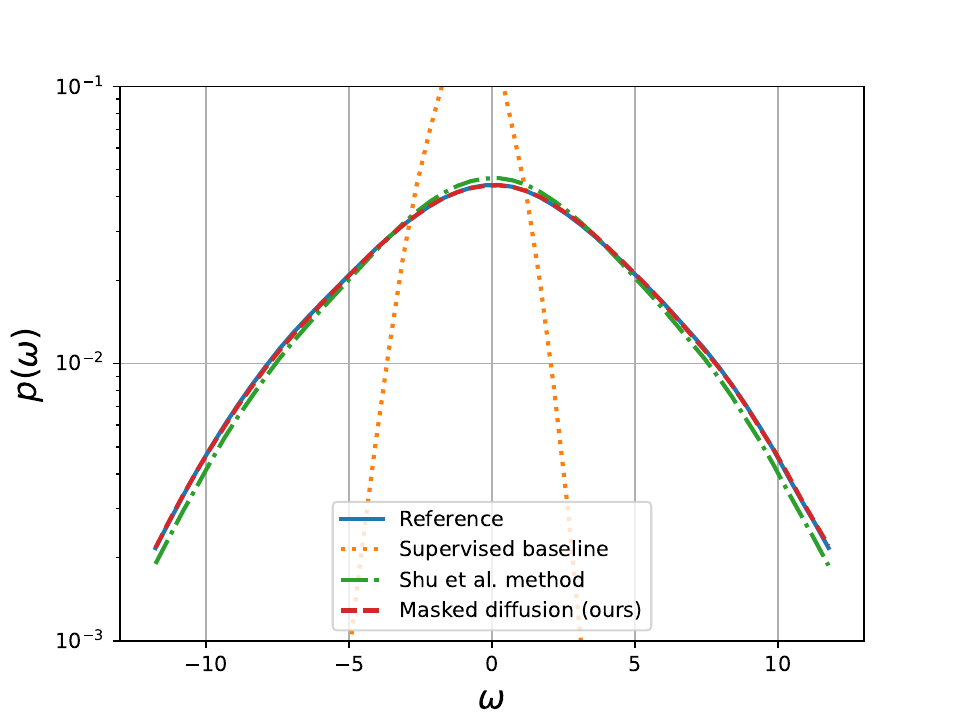}
        \caption{5\% sparse data reconstruction.}
        \label{fig:mag_rec_5}
\end{subfigure}
\begin{subfigure}[b]{0.49\linewidth}
    \includegraphics[width=0.85\linewidth]{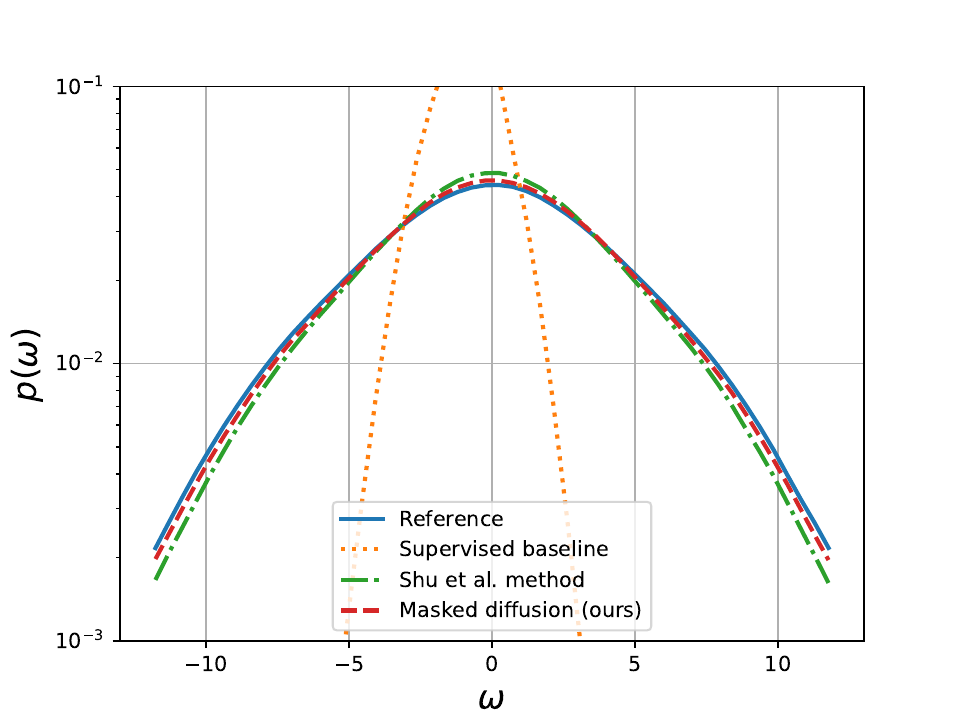}
        \caption{1.5625\% sparse data reconstruction.}
        \label{fig:mag_rec_1}
\end{subfigure}
\caption{2D Vorticity magnitude distribution of reconstructed samples.}
\label{fig:mag_rec}
\end{figure*}

We also study the distributional accuracy of the methods. For the 2D experiment, we measure the distribution of the vorticity magnitude of the reconstructed samples, as shown in Figure~\ref{fig:mag_rec}. Our method closely approximates the reference's vorticity distribution on the 5\% task, while Shu et al.'s method slightly underestimates the magnitude of the generated samples \cite{shu2023physics}. When evaluating the 1.5625\% task, our method also underestimates the vorticity, but gives a better distribution than Shu et al.'s method. 

\begin{figure*}[ht!]
\centering
\begin{subfigure}[b]{0.49\linewidth}
    \includegraphics[width=0.85\linewidth]{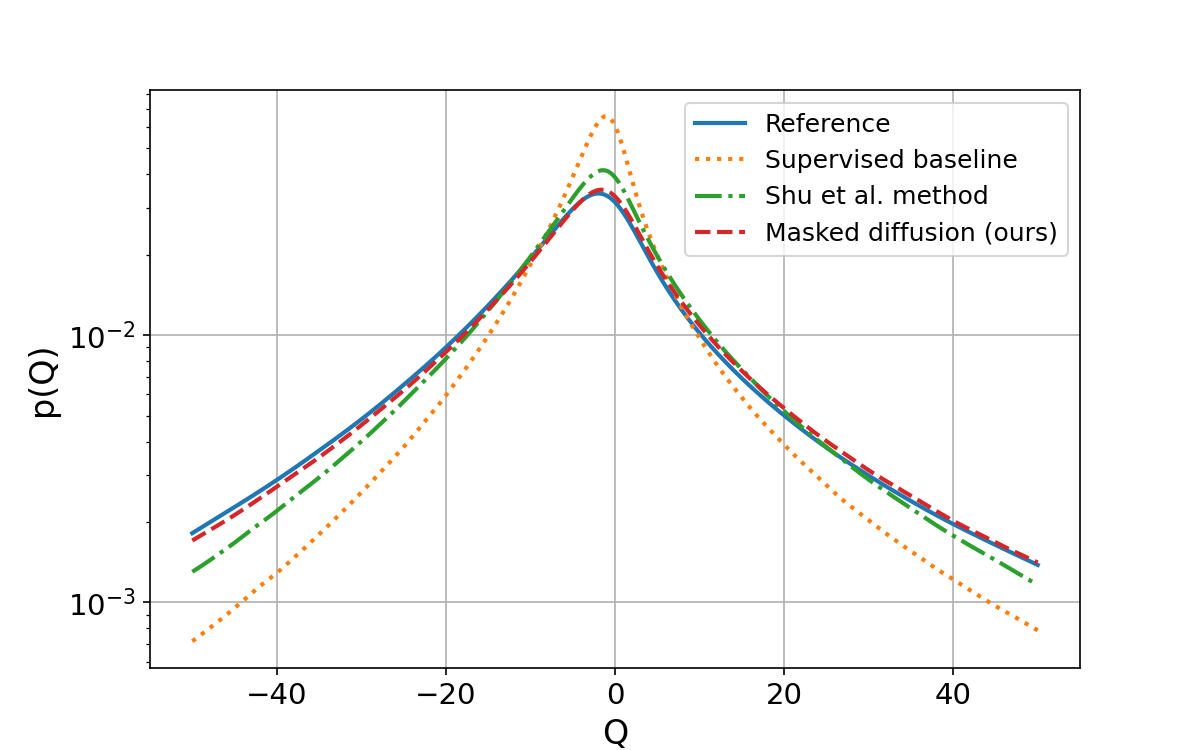}
        \caption{5\% sparse data reconstruction.}
\end{subfigure}
\begin{subfigure}[b]{0.49\linewidth}
    \includegraphics[width=0.85\linewidth]{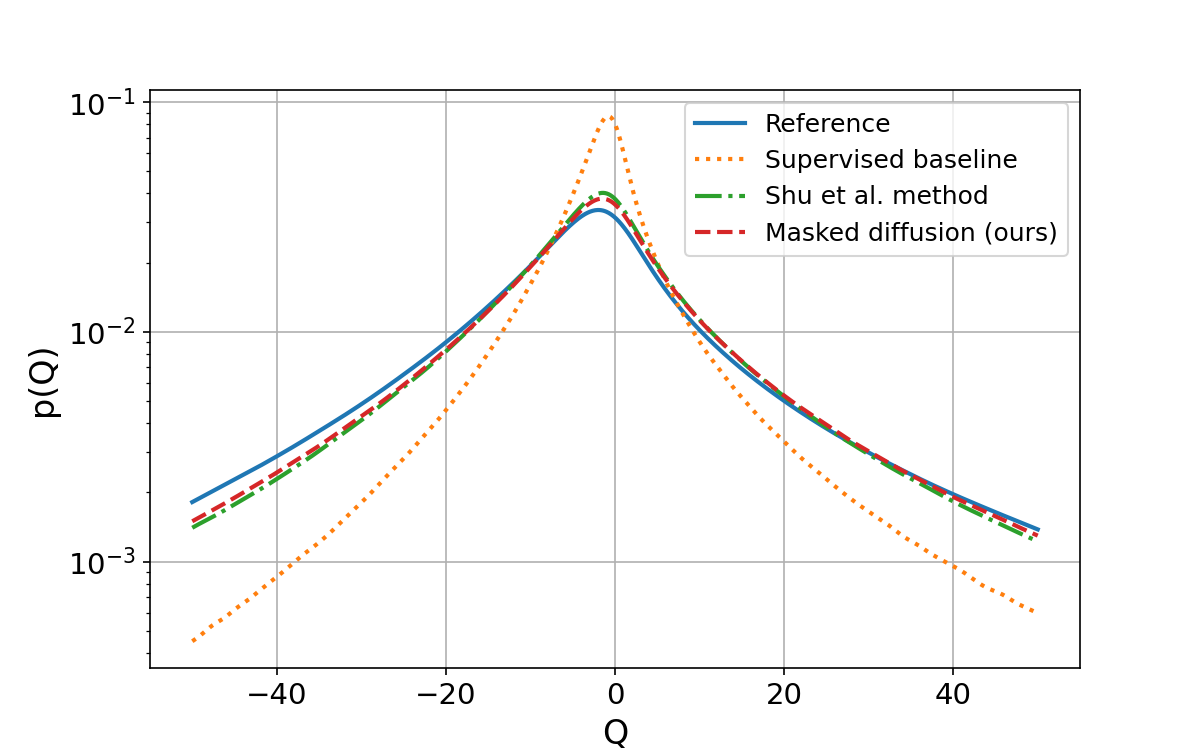}
        \caption{1.5625\% sparse data reconstruction.}
\end{subfigure}
\caption{Q-criterion distribution of reconstructed samples.}
\label{fig:q_dist}
\end{figure*}

\begin{figure*}[ht!]
  \centering
  \includegraphics[width=0.85\linewidth]{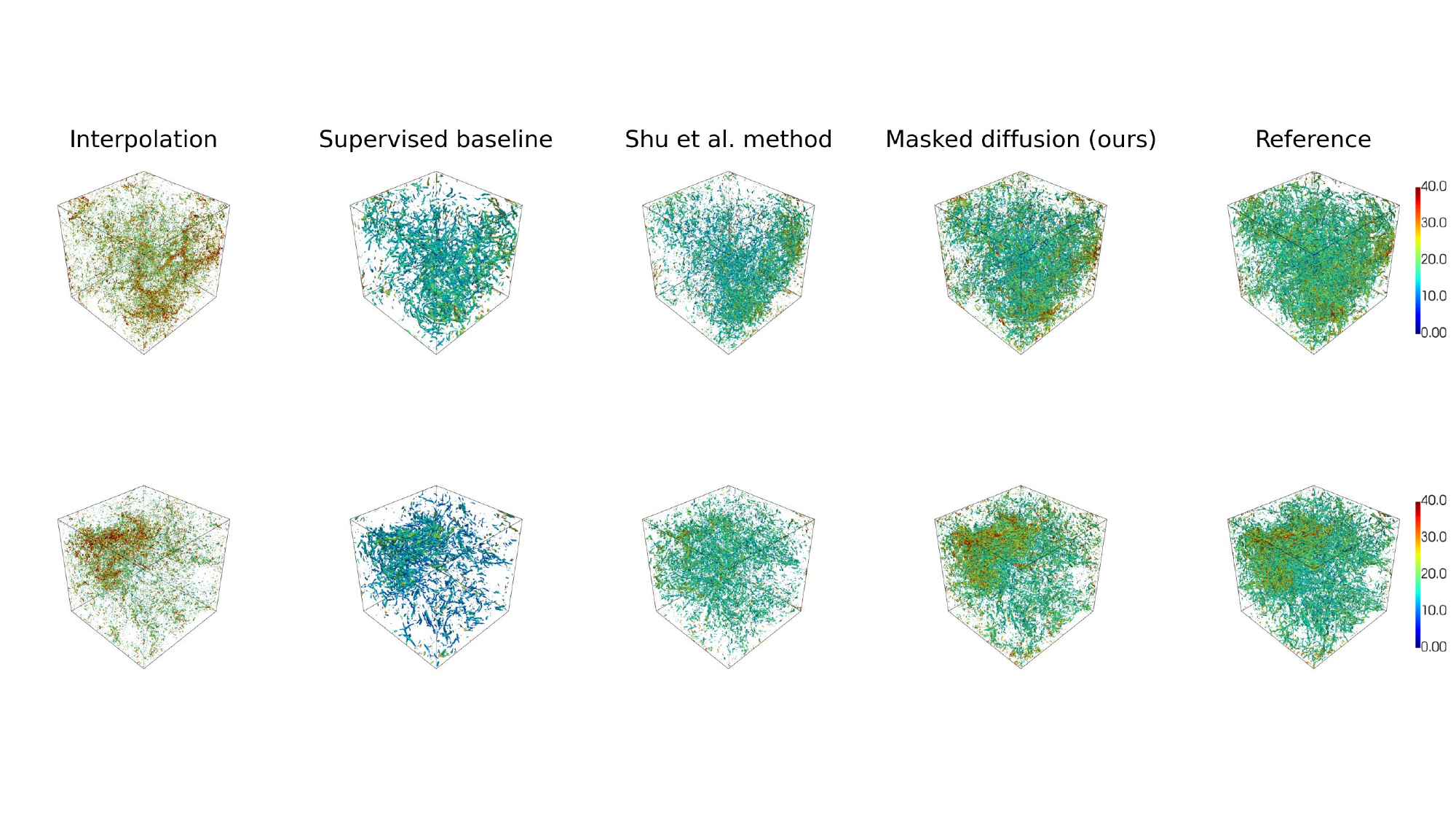}
  \caption{Q-citerion in 3D for the 5\% sparse data reconstruction. 3D plots showing Q isosurfaces (Q=80) with vorticity coloring.} 
  \label{fig:Q_reg5_iso}
\end{figure*}

For the 3D case, we compare the probability distribution of the Q-criterion, which is a scalar field used in fluid mechanics to identify vortices in a flow field. The definition of the $Q$-criterion is based on the decomposition of the velocity gradient into a symmetric part (strain rate tensor $S$) and an antisymmetric part (rotation tensor $\Omega$), and is computed as:
\begin{equation}
    Q = \tfrac{1}{2} \left( \|\Omega\|^2 - \|S\|^2 \right)
\end{equation}
\begin{equation}
S_{ij} = \tfrac{1}{2}\left(\frac{\partial u_i}{\partial x_j} + \frac{\partial u_j}{\partial x_i}\right), 
\hspace{0.2cm}
\Omega_{ij} = \tfrac{1}{2}\left(\frac{\partial u_i}{\partial x_j} - \frac{\partial u_j}{\partial x_i}\right)
\end{equation}
In Figure~\ref{fig:q_dist}, we clearly observe that our masked diffusion is the closest to the ground truth, while the supervised baseline overestimates the zero component. Figures~\ref{fig:Q_reg5_iso} show visualizations of the Q-criterion, highlighting how our proposed method is the only one that captures most of the smaller vortices in the fluid. 

When discussing efficiency, our method achieves these results using fewer reverse steps than Shu et al. \cite{shu2023physics}, making it a more computationally efficient model. For instance, for the more demanding 3D experiments in a NVIDIA L40S GPU, as shown in Appendix \ref{app:ablation reverse}, 100 reverse steps, i.e. NN evaluations, are enough to achieve maximum performance, compared to the T=1000 during training, or the 300 reverse steps used with Shu et al.'s method. This also results in shorter running times, where the supervised baseline requires around 0.04 s for a single sample, Shu et al.'s method takes 25 s, and our method takes 7.7 s. The only drawback of our method is a slight increase in memory demands. During masked diffusion, we need to store both the mask and the interpolated sparse data, resulting in 5.2 GB, compared with the 4.0 GB of Shu et al.'s method and the 1.4 GB of the single-step supervised baseline.

\subsection{Ablation study}
In this last section, we analyze the effect of different choices in our algorithm. First, the performance of diffusion models depends on the richness of our dataset \cite{super_res_2, super_res_3, training_data_1, training_data_2}. We now examine the relationship between the reconstruction error and the number of training samples for our 3D masked diffusion reconstructions in Table \ref{tab:training_data}. We present the metrics for models trained on the complete training dataset (100\%) and reduced versions with 50\% and 25\% of the samples, all during the same number of iterations. While the physics residual remains unaffected, reduced data amounts result in increased LSiM distances and spectral errors. Training with the complete dataset provides better results and less variability.

\begin{table*}[ht!]
\begin{ruledtabular}
    \centering
    \begin{tabular}{ l c c c c c}
    \textbf{Models} & \textbf{RMSE} & \textbf{P-RMSE} & \textbf{$\boldsymbol{L_\text{\textbf{res}}}$} & \textbf{LSiM} & \textbf{E.spectrum}\\
    \hline
    Masked diffusion (ours, default) 100\%  & \best{0.20 $\pm$ 0.01} & \best{0.18 $\pm$ 0.01} & \best{0.66 $\pm$ 0.03} & \best{0.09 $\pm$ 0.01} & \best{0.09 $\pm$ 0.01} \\

    Masked diffusion (ours, default) 50\%  & 0.22 $\pm$ 0.05 & \textbf{0.20 $\pm$ 0.08} & \best{0.66 $\pm$ 0.02} & \textbf{0.19 $\pm$ 0.16} & \textbf{0.21 $\pm$ 0.15}  \\

    Masked diffusion (ours, default) 25\%  & 0.22 $\pm$ 0.05 & 0.21 $\pm$ 0.06 & 0.71 $\pm$ 0.01 & 0.23 $\pm$ 0.17 & 0.24 $\pm$ 0.15 \\
    \end{tabular}
    \caption{Evaluation of the reconstruction task from 1.5625\% of the data for the 3D turbulent case. We compare our masked diffusion reconstructions with models trained using 25\%, 50\% and 100\% of the available training data.}
    \label{tab:training_data}    
\end{ruledtabular}
\end{table*}

\begin{figure}[ht!]
    \centering
    \includegraphics[width=0.45\textwidth]{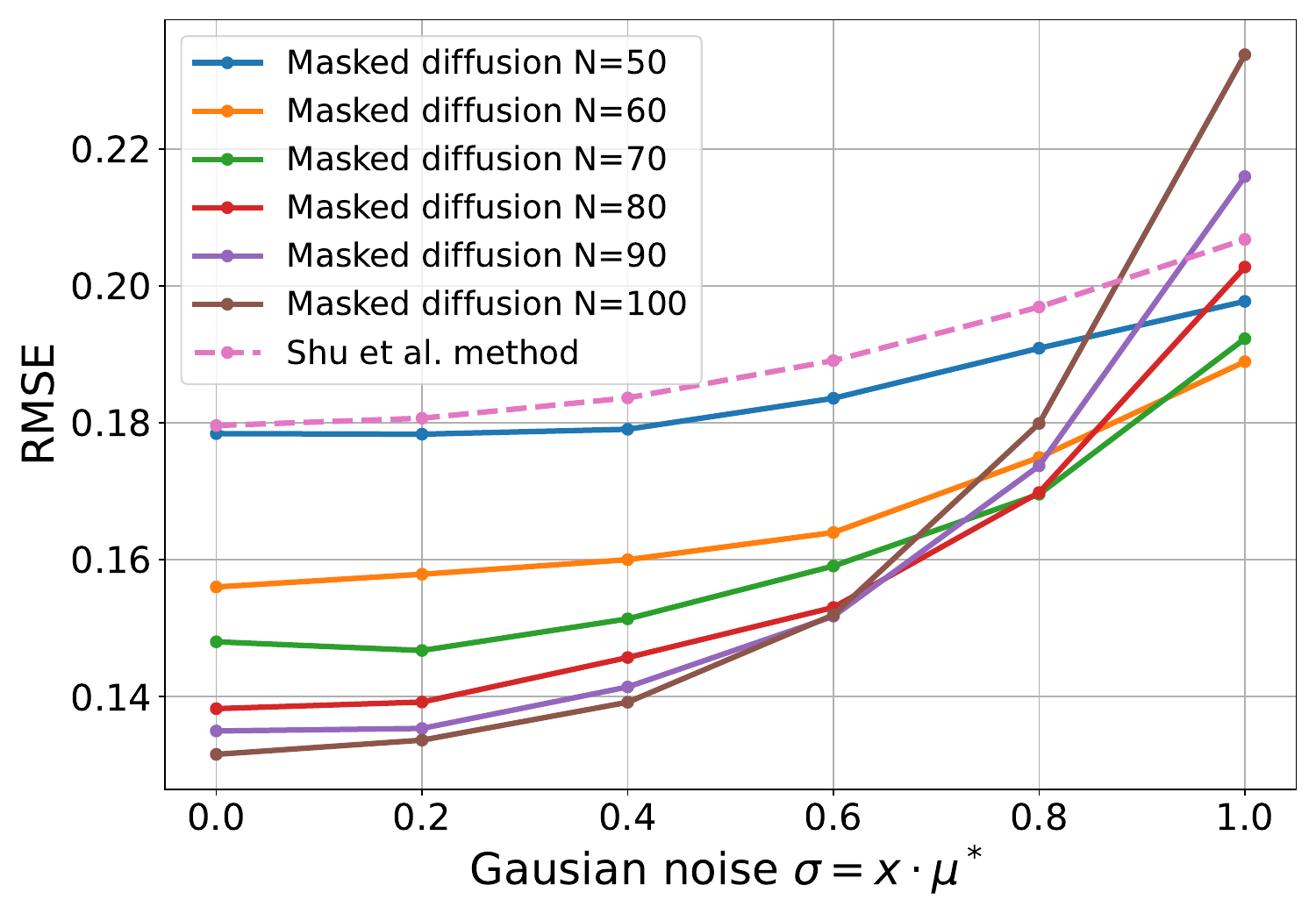}
    \caption{
    RMSE on the 5\% reconstruction task, where Gaussian noise $\sim\mathcal{N}(0, \sigma)$ is added to the input. The $\sigma$ is chosen to be a percentage of the dataset's mean $\mu^*$.
    }
    \label{fig:noise_robustness}
\end{figure}

Finally, we study the robustness of our method to noisy inputs. We evaluate the RMSE of our model on the 5\% reconstruction task when Gaussian noise is added to the input. The model achieves its best accuracy when performing 100 neural network evaluations (as discussed in Appendix \ref{app:ablation reverse}) with noiseless input. However, in the presence of noise, we see a rapid decrease in accuracy when performing 100 reverse steps. We observe an inverse relationship between the ideal number of neural network evaluations and the amount of noise. 
In Figure \ref{fig:noise_robustness}, one can see how, when working with noisy inputs, the performance increases as the number of neural network evaluations decreases. Thus, the number of denoising steps can be interpreted as the strength of the guidance, which should be stronger when we are more certain about the precision of the input, and weaker when it is unreliable. Figure \ref{fig:noise_robustness} shows that the number of reverse steps to perform a reconstruction should be chosen depending on the amount of noise that is present in the input. 

\begin{figure}[ht!]
    \centering
    \includegraphics[width=0.43\textwidth]{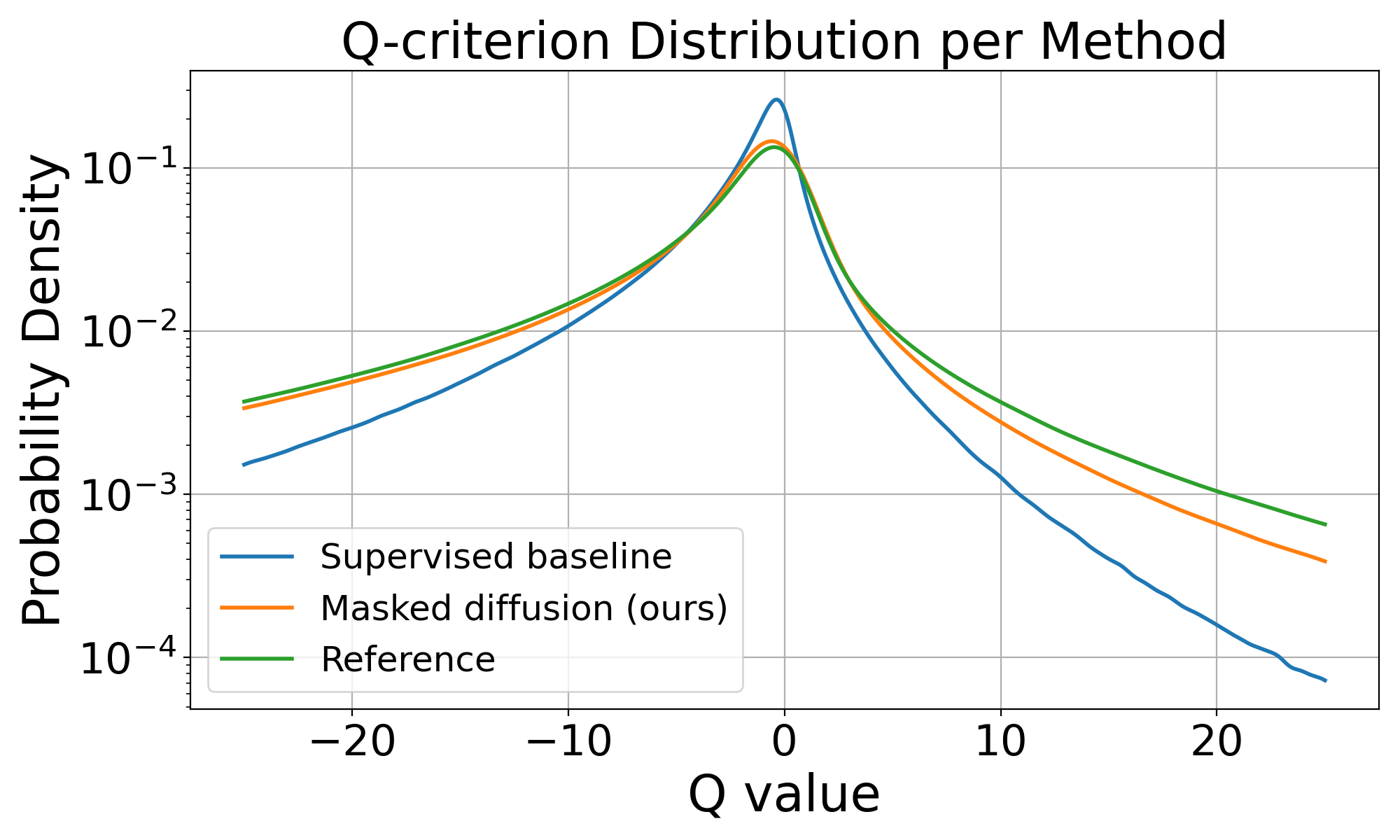}
    \caption{
    Q-criterion distribution for the reconstructions of 3D TCF data with a 0.1\% sparsity.
    }
    \label{fig:ch_q}
\end{figure}

\begin{figure}[ht!]
    \centering
    \includegraphics[width=0.43\textwidth]{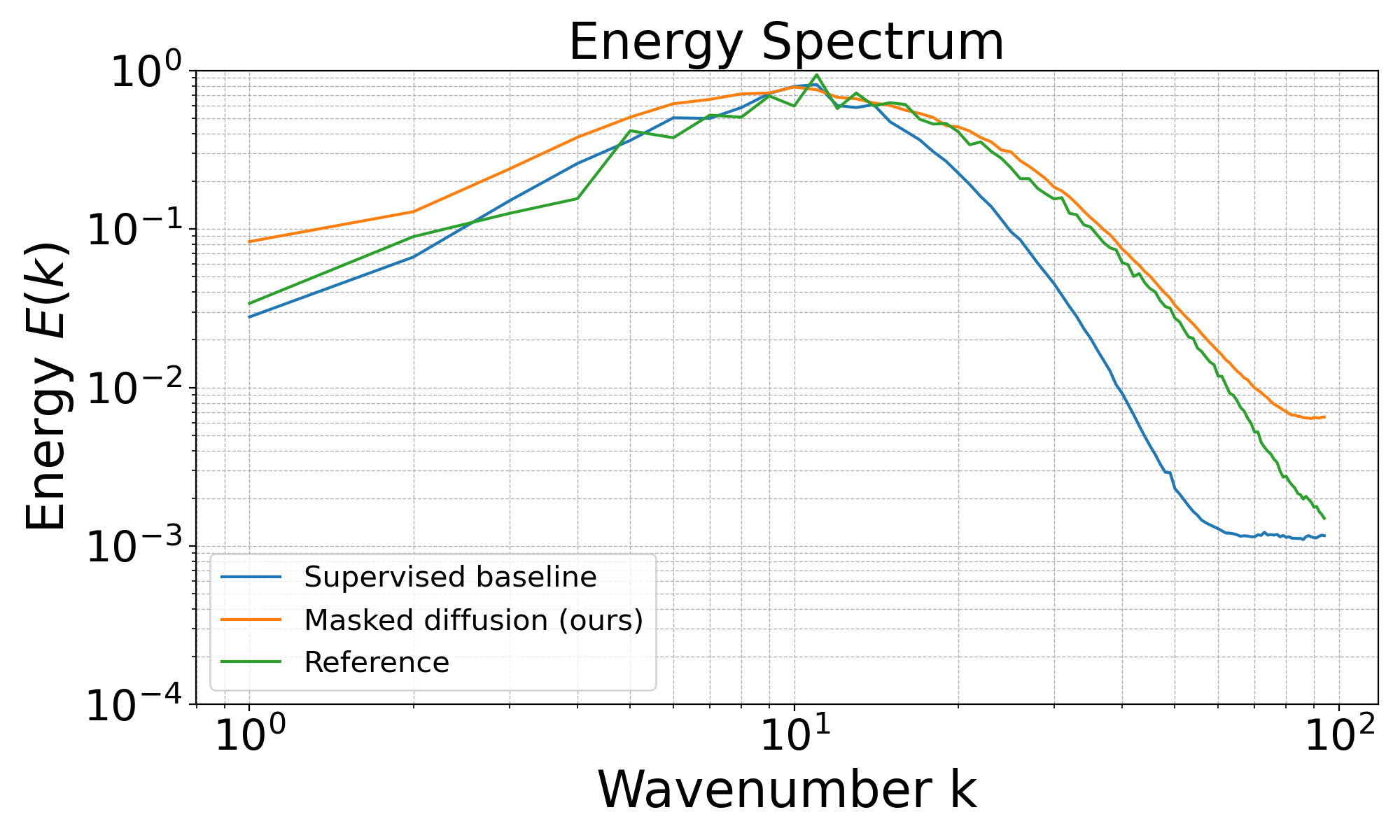}
    \caption{
    Energy spectrum for the reconstructions of 3D TCF data with 0.1\% sparsity.
    }
    \label{fig:ch_freq}
\end{figure}

\subsection{Summary of experiments and 3D outlook}

Overall, our proposed training and sampling techniques demonstrate a clearly improved performance compared to existing approaches. These results highlight the ability of our masked diffusion framework to capture fine-grained details and preserve the structural integrity of the reconstructed fluid fields.
To provide an outlook towards further 3D applications of our approach, we investigated the performance of our method on 3D turbulent channel flow (TCF) data at friction Reynolds number $Re_{\tau} = 180$~\cite{guastoni_channel}, obtaining results that align with the results reported above. 
We observed that the supervised baseline can achieve almost perfect reconstructions with  5\% and 1.5\% sparse data. However, the reconstruction performance degrades when less data is available (0.1\%). In this case, our masked diffusion method performs much better, reconstructing most of the important turbulence scales, as shown in Figures \ref{fig:ch_q} and \ref{fig:ch_freq}.

\section{Conclusions}

In this work, we proposed a novel method for reconstructing high-fidelity flow fields from limited sparse data using diffusion models. Instead of generating a random sample of the target distribution, our technique utilises the available sparse information to guide the sampling process towards our desired reconstructed sample. This results in trained models that better approximate the fluid's structure, as they generate the missing information utilising the available data as a reference for the overall structure. Moreover, we introduced the $\text{ConFIG}_\text{u}$ training method to ensure physically accurate solutions, and showed how multiple targets can be combined in a non-destructive manner.

The central idea of masked diffusion can be applied to super-resolution problems \cite{super_res_1, super_res_2, super_res_3}, as these tasks are equivalent to sparse data reconstruction where the available data has a particular Cartesian structure, instead of being randomly placed. Applications additionally extend to experimental fluid mechanics, where a sparsely distributed set of probes can provide measurements at a limited number of locations, and the objective is to reconstruct the entire flow field. In such setups, obtaining complete flow field information is fundamentally constrained by the limitations of measurement techniques. Deploying a large number of probes is often infeasible, as intrusive sensors such as hot-wire anemometers, Pitot tubes, or pressure taps can disturb the flow and bias the measurements. Even advanced non-intrusive methods, such as Particle Image Velocimetry (PIV), are subject to trade-offs between field of view, resolution, and optical accessibility, resulting in data that is inherently sparse and noisy. In comparison to supervised models~\cite{guastoni2025fully}, the use of a generative approach can improve the reconstruction of the small scales of the flow field.

We have shown that our method outperforms previous diffusion-based reconstruction approaches in both 2D and 3D scenarios. In particular, it excels at producing physically consistent reconstructions while preserving key flow statistics and accurately capturing the smallest scales. Despite that, it is subject to certain limitations inherent to any reconstruction approach, such as the need for specific data to train the underlying network. Moreover, in scenarios where frequent retraining of the model is necessary, our training method $\text{ConFIG}_u$ may not be ideal, as it is computationally more expensive than standard diffusion training. On the other hand, our approach outperforms previous diffusion-based works at inference time, as fewer reverse steps are needed.

Our results point to several interesting avenues for future work, such as evaluating how the proposed generative method performs for reconstructing other flow problems. 
We believe that our efforts demonstrate the promise of diffusion models for highly accurate reconstructions of physical flow fields given sparse measurements, an important open problem in different fields of engineering. Our work also highlights the importance of non-conflicting algorithms when training generative models with multiple objectives and physics constraints. Our approach generally encourages further research into physics-constrained diffusion models for addressing probabilistic flow-related problems.

\begin{acknowledgments}
This work was partially supported by the ERC Consolidator Grant \textit{SpaTe} (CoG-2019-863850) and by the China Scholarship Council (No. 202206120036). 

\end{acknowledgments}

\section*{Data Availability Statement}
The data and source code supporting the findings of this study are openly available on GitHub at \url{https://github.com/tum-pbs/super-resolution}.

\section*{Author Declarations}
\subsection*{Author Declarations}
The authors have no conflicts to disclose.

\section*{Author contributions}

\textbf{Marc Amorós-Trepat}: Conceptualization (equal); Methodology (equal); Investigation (equal); Data Curation (equal); Formal Analysis (equal); Software (lead); Validation (equal); Visualization (equal); Writing – Original Draft (equal); Writing – Review \& Editing (equal);
\textbf{Luis Medrano-Navarro}: Investigation (equal); Data Curation (equal); Formal Analysis (equal); Software (equal); Validation (equal); Visualization (equal); Writing – Original Draft (equal); Writing – Review \& Editing (equal);
\textbf{Qiang Liu}: Conceptualization (equal); Methodology (equal); Writing – Original Draft (equal); Writing – Review \& Editing (equal); Supervision (equal);
\textbf{Luca Guastoni}: Writing – Review \& Editing (equal); Supervision (equal);
\textbf{Nils Thuerey}: Writing – Review \& Editing (equal); Supervision (equal);

\appendix

\section{Comparison between point-wise and Gaussian masking methods}
\label{app:masks}

In Section~\ref{sec:meth_sampling}, we introduced a masking strategy to improve the prediction of the resulting high-fidelity field at any given timestep $t$. We proposed two variants of this mask: one that activates only at specific points where the ground truth value is known, and another that incorporates a Gaussian smoothing function around each known value. Table~\ref{tab:masks_comp} compares these methods with the task of reconstructing a flow field from 1.5625\% of the data, showing how the Gaussian mask consistently outperforms the point-wise mask across all metrics introduced in the Results section. 
\begin{ruledtabular}
\begin{table}[ht!]
    \centering
    \begin{tabular}{l c c c c}
        \textbf{Methods} & \textbf{RMSE} & \textbf{$\Vert \mathcal{R}(\tilde{\omega}) \Vert_2$} & $\boldsymbol{L_\text{\textbf{res}}}$ & \textbf{LSiM} \\ 
        \hline
        Interpolation & 0.4930 & 106.2809 & 102.9909 & \textbf{0.1552} \\ 
        Point-wise Mask & \textbf{0.3500} & \textbf{4.2134} & \textbf{0.9234} & 0.1645 \\ 
        Gaussian Mask & \best{0.3024} & \best{3.5658} & \best{0.2758} & \best{0.1094} \\ 
    \end{tabular}
    \caption{Comparison between nearest interpolation and our proposed masking methods. The evaluated task is the reconstruction of a field from 1.5625\% of the known values. The used model was trained with the $\text{ConFIG}_u$ method.}
    \label{tab:masks_comp}
\end{table}
\end{ruledtabular}

\section{Training details}
\label{sec:hyper-params}

The 2D network is based on a UNet model \cite{ronneberger2015u}, which is a convolutional neural network architecture that features a symmetric encoder–decoder structure with skip connections to capture both global context and fine details. To further expand the receptive field of the network, we also utilize self-attention \cite{vaswani2017attention} in the bottleneck layer.

As a backbone model for 3D fields, we use the P3D model \cite{holzschuh2025p3dscalableneuralsurrogates}, a recent hybrid CNN-Transformer architecture for 3D physical simulations that can be applied in patches or to the whole sample. It follows a typical hierarchical U-shaped structure, comprising an encoder and a decoder, based on a combination of convolutional and transformer blocks. These blocks combine ideas from Swin Transformers \cite{swinir} and diffusion transformers (DiT) \cite{diff_trans}.

To obtain comparable results, all 2D models were trained for 1000 epochs and a batch size of 5. A learning rate of $10^{-4}$ was used, with an exponential scheduler with $\gamma=0.98$, and a minimum $lr=10^{-5}$. The training of the 3D models was slightly different. Due to having a smaller number of samples, we trained for 2000 epochs to achieve convergence. Since the samples are much larger, we needed a more powerful GPU (NVIDIA L40S with 50 GB) for training and a slightly different batch size (4). We still used the same variable learning rate. The noise schedule $\beta$ for the diffusion forward process was kept as linear between $10^{-4}$ and 0.02.

\section{Ablation of interpolation methods for guidance in masked diffusion}\label{app:interpolation}
Our masked diffusion method utilizes an interpolation of the sparse data to guide the reverse process. For simplicity, our first choice was to use nearest interpolation for this purpose. Nonetheless, one could use more advanced interpolation techniques. For example, we also considered linear interpolation and present a comparison on the 3D dataset in Tables \ref{tab:task7_app} and \ref{tab:task8_app}.
We observe slightly better results in terms of RMSE and residual error for linear interpolation, but a worse energy spectrum error. Additionally, computing a more advanced interpolation also requires extra runtime, which could be even higher than the runtime of the diffusion sampler itself. Therefore, we generally prefer to keep the nearest interpolation.

\begin{table*}[ht!]
\begin{ruledtabular}

    \centering
    \begin{tabular}{ l c c c c c}
    \textbf{Models} & \textbf{RMSE} & \textbf{P-RMSE} & \textbf{$\boldsymbol{L_\text{\textbf{res}}}$} & \textbf{LSiM} & \textbf{E.spectrum}\\
    \hline
    Masked diffusion "nearest" (ours, default) & 0.17 $\pm$ 0.01 & \best{0.11 $\pm$ 0.01} & 0.61 $\pm$ 0.03 & 0.06 $\pm$ 0.01 & \best{0.05 $\pm$ 0.01} \\
    Masked diffusion "linear" (ours) & \best{0.16 $\pm$ 0.01} & 0.20 $\pm$ 0.01 & \best{0.50 $\pm$ 0.02} & 0.06 $\pm$ 0.01 & 0.16 $\pm$ 0.01 \\
    \end{tabular}
    \caption{Evaluation of the reconstruction task from 5\% of the data for the 3D turbulent case.}
    \label{tab:task7_app}    
\end{ruledtabular}
\end{table*}

\begin{table*}[ht!]
\begin{ruledtabular}
    \centering
    \begin{tabular}{ l c c c c c}
    \textbf{Models} & \textbf{RMSE} & \textbf{P-RMSE} & \textbf{$\boldsymbol{L_\text{\textbf{res}}}$} & \textbf{LSiM} & \textbf{E.spectrum}\\
    \hline
    Masked diffusion "nearest" (ours, default) & 0.20 $\pm$ 0.01 & \best{0.18 $\pm$ 0.01} &  0.66 $\pm$ 0.03 & 0.09 $\pm$ 0.01 & \best{0.09 $\pm$ 0.01} \\
    Masked diffusion "linear" (ours) & \best{0.19 $\pm$ 0.01} & 0.21 $\pm$ 0.01 & \best{0.53 $\pm$ 0.02} & 0.09 $\pm$ 0.01 & 0.18 $\pm$ 0.01 \\
    \end{tabular}
    \caption{Evaluation of the reconstruction task from 1.5625\% of the data for the 3D turbulent case.}
    \label{tab:task8_app}    
\end{ruledtabular}
\end{table*}

\section{Very sparse reconstruction with 0.1\% input data }\label{app:extreme}
\begin{figure*}[ht!]
  \centering
  \includegraphics[width=0.85\textwidth]{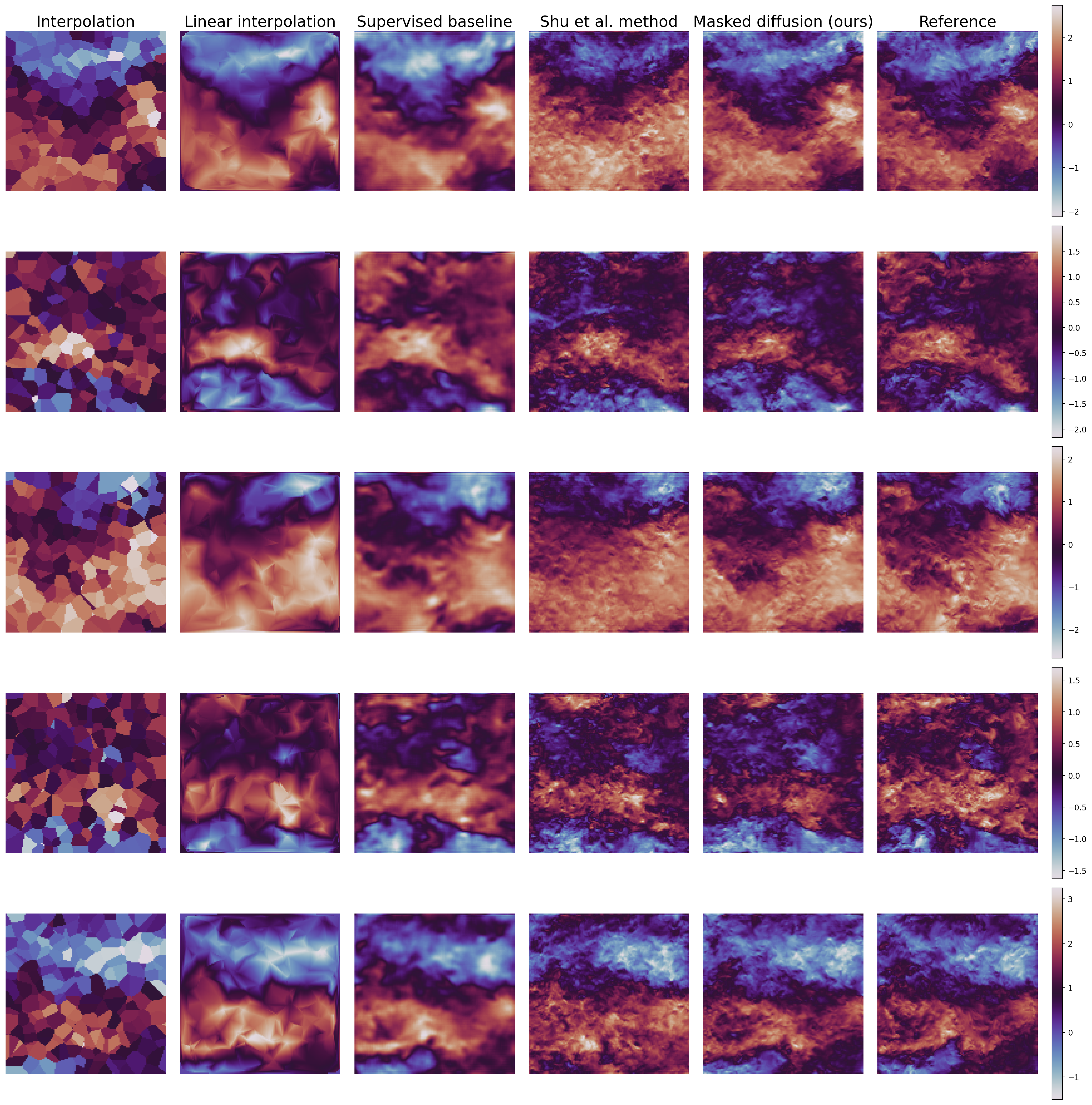}
  \caption{Example 3D isotropic turbulence 0.1\% sparse data reconstruction (2D slices).}
  \label{fig:reg_point1}
\end{figure*}

To support the results in Table \ref{tab:task_point1} and Figure \ref{fig:rmse_spectrum_point1} with qualitative examples, we include visual reconstruction results in Figure \ref{fig:reg_point1}. These visualizations show that the masked diffusion method provides plausible reconstructions despite the extremely sparse inputs, thanks to its ability to recreate the smallest structures in the fluid.

\section{Ablation of DDIM reverse steps}
\label{app:ablation reverse}
\begin{figure}[ht!]
  \centering
  \includegraphics[width=\linewidth]{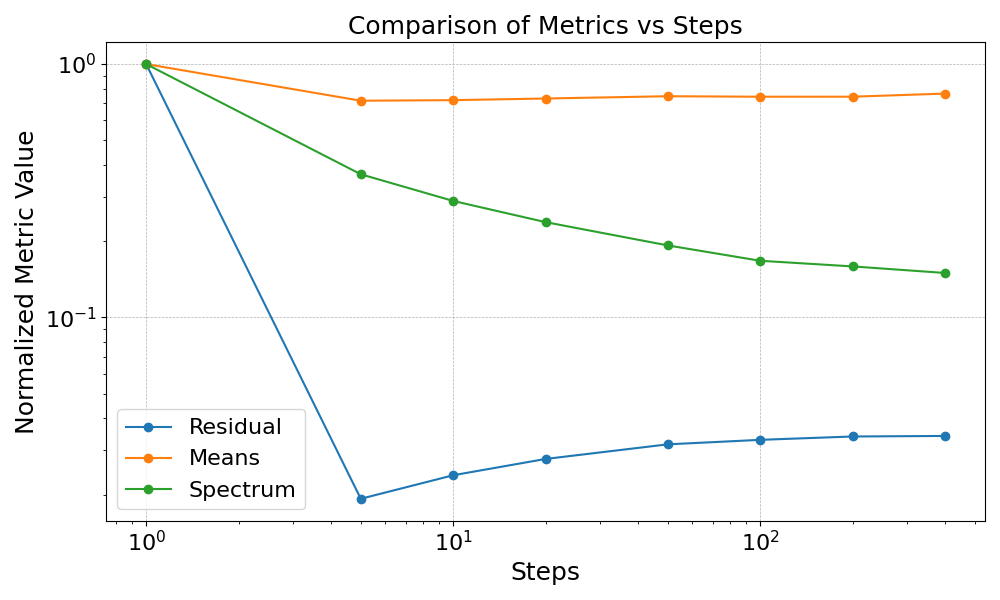}
  \caption{Optimization of DDIM reverse steps. Evolution of three normalized metrics for an increasing number of reverse steps.} 
  \label{fig:3d_metrics_steps}
\end{figure}

We optimized the choice of the number of reverse steps in DDIM. In Figure \ref{fig:3d_metrics_steps}, for the 3D case, the metrics stabilize after around 100 steps. A similar result is obtained in the 2D case. Thus, we chose this value for the final experiments.

\bibliography{references}

\end{document}